\documentclass[aps,pra,twocolumn, showpacs,showkeys,secnumarabic,amssymb,amsmath,superscriptaddress]{revtex4-2}
\usepackage[utf8]{inputenc}
\usepackage{amssymb,amsthm}
\usepackage{bbold}
\usepackage{color}
\usepackage{graphics}
\usepackage{graphicx}
\usepackage{enumitem}
\usepackage{bm}
\usepackage{mathtools}
\usepackage{hyperref}
\usepackage{appendix}
\usepackage{xpatch} 
\usepackage{comment}

\usepackage{CJK}
\usepackage{indentfirst}

\begin{document}
\title{The stabilizer ground state and applications to quantum simulation}

\author{Yuping Mao}
\affiliation{School of Science, Huzhou Normal University, Huzhou, Zhejiang 313000, China}

\author{Jiaxing Feng}
\affiliation{School of Science, Huzhou Normal University, Huzhou, Zhejiang 313000, China}

\author{Yimeng Mao}
\affiliation{School of Science, Huzhou Normal University, Huzhou, Zhejiang 313000, China}

\author{Tim Byrnes}
\email{tim.byrnes@nyu.edu}
\affiliation{Department of Physics, New York University, New York, NY, 10003, USA}
\affiliation{Center for Quantum and Topological Systems (CQTS),
NYUAD Research Institute, New York University Abu Dhabi, UAE}


\date{November 2026}
\begin{abstract}
The stabilizer ground state is the lowest energy stabilizer state with respect to a given Hamiltonian.  This is useful in quantum simulation contexts as it allows for a Clifford circuit approximation of a ground state that can be further refined towards the true ground state. However, in many cases it is highly degenerate and does not give a unique stabilizer state.  In this paper, we define and investigate methods to prepare the optimal stabilizer ground state (OSGS), which is the stabilizer ground state that has the highest fidelity with the true ground state.  
We show how the optimal stabilizer ground state may be evaluated by classically selecting the optimal stabilizer group and 
employing the stabilizer tableaux formalism. 
We show applications of the OSGS in the context of measurement-based deterministic imaginary time evolution, where it is used as a highly efficient initial state for the evolution towards the true ground state.  
Candidate commuting subsets are generated from the target Hamiltonian using a genetic search guided by energy evaluations incorporating an inner optimization over generator signs. 
Once a complete signed generating set is specified, the corresponding state can be efficiently prepared using Clifford gates. 
Complexity analysis further shows that the algorithm's  resource scales polynomially with system dimension, highlighting its efficiency and potential quantum advantage.
\end{abstract}
\maketitle

\section{Introduction}
In recent years, quantum computing has experienced rapid growth with impressive improvements in quantum hardware \cite{gyongyosi2019survey,ladd2010quantum,steane1998quantum,williams2010explorations,bluvstein2024logical,arute2019quantum}. A broad range of quantum applications are gradually moving towards practical implementation, particularly in the field of quantum simulation \cite{georgescu2014quantum,nielsen2010quantum,bloch2008many,byrnes2021quantum,daley2022practical}.  
The primary objective of quantum simulation is to reproduce the behavior of physical systems using controllable quantum systems. This approach has a variety of applications in many fields, such as quantum chemistry, high-energy physics, and condensed matter physics \cite{levine2009quantum, bauer2023quantum, nachman2021quantum,hofstetter2018quantum,byrnes2006simulating,byrnes2007quantum,dovesi2018quantum, fauseweh2024quantum}. 
A common feature among these applications is that the target problems involve quantum many-body systems, which can be encoded onto a quantum computer \cite{tasaki2020physics,vidal2004efficient,eisert2015quantum}. Compared with classical computation, quantum simulation has the potential to exhibit significantly higher efficiency in solving such problems \cite{lloyd1996universal}.  

Quantum state preparation plays an essential role in both quantum simulation \cite{abrams1997simulation,wang2009efficient} 
and other related methods \cite{aharonov2003adiabatic,aspuru2005simulated}. 
For example, initial state preparation is an important step in quantum simulation methods using the phase estimation algorithm \cite{cleve1998quantum}. Here, the phase estimation algorithm is used to measure and project the quantum state in the energy eigenbasis of a Hamiltonian.  The success probability of obtaining the ground state, for instance,  is equal to the fidelity of initial state with respect to the ground state.  As such, preparing an initial state with minimal error can significantly improve both efficiency and accuracy of quantum algorithms \cite{long2001efficient, ge2019faster,gleinig2021efficient,berry2025rapid}. 
Consequently, quantum state preparation remains one of the central tasks in quantum simulation and quantum computation \cite{georgescu2014quantum, poulin2009preparing}.
More recently, several other quantum simulation methods have been proposed, including variational quantum eigensolver (VQE)-based approaches \cite{kandala2017hardware, cerezo2022variational}, quantum compressed sensing-based encoding \cite{gross2010quantum,flammia2012quantum,kueng2017low}, and random projection methods  \cite{stannigel2014constrained}. 
For some particular class of systems, numerous specialized state preparation protocols have been developed, such as those for fermionic systems \cite{abrams1997simulation, tarruell2018quantum, hensgens2017quantum}, molecular systems \cite{wang2009efficient, huggins2025efficient}, and lattice models \cite{chiu2018quantum, muller2007state}. 
While these algorithms provide broadly applicable state preparation strategies, they are typically resource-demanding, posing significant challenges for noisy intermediate-scale quantum (NISQ) devices.

To address this challenge, the stabilizer ground state (SGS) --- defined as the stabilizer state that minimizes the stabilizer group energy  --- has attracted recent attention. 
Sun, Cheng, and Zhang established a framework for determining stabilizer ground states of Pauli Hamiltonians and developed an exact algorithm for one dimensional systems with fixed-range local interactions   \cite{sun2025stabilizer}.
Stabilizer states play a central role in quantum error correction and fault-tolerant quantum computation \cite{laflamme1996perfect,kitaev1997quantum,gottesman2009introduction}.  
The advantages of stabilizer ground states include their efficient classical description and compatibility with fault-tolerant quantum operations. 
In the Lipkin-Meshkov-Glick model, Robin demonstrated that stabilizer reference states can capture collective entanglement and accelerate imaginary time propogation when they provide a larger overlap with the exact ground state \cite{robin2025stabilizer}.
This allows SGS to be excellent candidates for initial states of various quantum simulation methods.  One issue with SGS is that there is no guarantee that they are non-degenerate.  In fact, in practice, the SGS have a high degeneracy and fidelities with the ground state are rather variable.  This leaves some room for improvement in the utilization in various quantum simulation algorithms, as often what is required is a high fidelity initial state, rather than merely a low energy state.  

In this paper, we investigate efficient ground-state preparation methods for quantum simulation. 
We define the {\it optimal stabilizer ground state} (OSGS) as the SGS that has the highest fidelity with the ground state.
This state provides a suitable starting point for subsequent ground-state refinement in quantum simulation.
We discuss how the OSGS can be determined from a given Hamiltonian, without performing numerically prohibitive exhaustive search. 
Within the stabilizer formalism underlying the Gottesman-Knill theorem, its preparation circuit can be efficiently constructed classically and implemented using Clifford gates.
We then demonstrate its suitability as an initial state for further refinement toward the true actual ground state using the measurement-based imaginary time evolution (MITE) algorithm.  MITE is a deterministic algorithm for preparing ground states through a sequence of measurements and feed-forward operations.  
Due to the high fidelity of the OSGS with the ground state, this substantially reduces the required depth of the MITE algorithm.

This paper is organized as follows. In Sec. \ref{sec2}, we briefly review the stabilizer formalism and introduce the definitions and methods we employed in this work. In Sec. \ref{sec3}, we introduce our algorithm to find the OSGS. In Sec. \ref{sec4}, we give further analysis of our algorithm to prepare the OSGS, giving a concrete example of the procedure for the transverse-field Ising model. 
In Sec. \ref{sec5}, we describe how the OSGS state preparation can be applied to the MITE scheme, analyze the computational complexity and convergence error.  In Sec. \ref{sec:val} we present numerical simulations for the OSGS initilized MITE procedure.   
Sec. \ref{sec6}, summarizes the main results.

\section{ Basic Definitions} \label{sec2}

In this section, we review concepts such as the stabilizer formalism and magic measures, then define the SGS and OSGS.  

\subsection{Stabilizer formalism}

The stabilizer formalism is a framework for characterizing quantum states, originally introduced by Gottesman \cite{gottesman1996class,gottesman1997stabilizer}. 
For an $N$-qubit system, a stabilizer state $|S \rangle$ is the unique simultaneous $+1$ eigenstate, up to a global phase, of the elements of its full stabilizer group \cite{sarkar2021sets}
\begin{align}
      &G =\{s_iP_1,s_2P_2,\dots,s_{2^N}P_{2^N} \},  \label{stabgroup} \\
    &s_iP_i|S\rangle   =|S\rangle.
\end{align}
Here, sign $s_i\in\{-1,+1\}$ and the unsigned Pauli strings $ P_i \in {\cal P} $ are tensor products of Pauli operators.  The set of all Pauli strings for $N $ qubits is
\begin{align}
&\mathcal{P}=\{ I, X, Y, Z\}^{\bigotimes N}.
\end{align}
There are a total of $|\mathcal{P}|=4^{N}$ Pauli strings.  Every generator in group $G$ can be taken as the Pauli string $ P_i $ multiplied by the sign $s_i$.  

An Abelian subgroup $G$ of $N$ qubit Pauli group containing $2^N$ elements, excluding $-I$, specifies a unique stabilizer state as its common $+1$ eigenstate.
An independent generating set $g$ is a subset of $G$ whose elements generate all elements of $G$ through matrix multiplication.
It can be written as
\begin{align}
    &g = \{s_1P_1, s_2P_2, \dots, s_N P_N \},
    \label{generators}
\end{align}
where $|g|=N$, $s_j\in\{-1,+1\}$, and $P_j \in \mathcal{P}$.
We note that every generator in $g$ satisfies $g_j=s_jP_j$.
Here, independence refers to linear independence of the binary Pauli representations over $\mathbb{F}_2$.
We use $g$ to denote the independent generating set, and $G=\langle g\rangle$ to denote the full stabilizer group.

The stabilizer state $|S_g\rangle$ is uniquely determined by $g$ up to a global phase, with its projector being written
\begin{equation}
|S_g\rangle\langle S_g|= \prod_{j=1}^N\frac{I+s_jP_j}{2}, 
\end{equation}
where $ s_jP_j\in g $.  More generally, a set of $r$ independent commuting stabilizer generators defines a common $+1$ eigenspace of dimension 
\begin{equation}
    \text{dim}({\cal H})=2^{N-r}.
\end{equation}
Thus, $N$ independent generators are required to specify a unique pure stabilizer state.

Let ${\cal S} = \{ |S_1 \rangle, |S_2 \rangle, \dots, |S_{D_S}\rangle  \}$ denote the set of all pure $N$-qubit stabilizer states, counted up to global phase.  
A convex mixture of pure stabilizer states, referred to here a mixed stabilizer state, can be written as 
\begin{equation}
 \rho_S = \sum_{j=1}^{D_S} p_j  |S_j \rangle \langle S_j |,    
\end{equation}
where $ p_j \geq 0 $ and $ \sum_j p_j =  1 $, so that $\{p_j\}$ forms a probability distribution. 
The total number of pure stabilizer states is
\begin{align}
 D_S =    2^N \!\prod_{n=1}^N (2^n+1) .
\end{align}
This number grows superexponentially with $N$, as $D_s=2^{N^2/2+O(N)}$.

The stabilizer formalism provides an efficient framework for classically simulating Clifford circuits acting on stabilizer states, as established by the Gottesman-Knill theorem \cite{gottesman1998heisenberg,ermakov2026unified}.  Although stabilizer states can be highly entangled, they do not encompass all quantum states obtainable through arbitrary unitary operations.
The Clifford gate set can be extended to a universal gate set by adding a suitable non-Clifford gate, such as the $T$ gate.

\subsection{Symplectic representation}\label{subsla}
In symplectic linear algebra, a Pauli string can be represented by a binary vector \cite{crawford2021efficient}:
\begin{align}
    &P_i\rightarrow(x_i|z_i)\in \mathbb{F}_2^{2N}.
\end{align}
For example, for a single qubit, the corresponding entries are assigned according to 
\begin{equation}
    I\leftrightarrow (0|0),\ X\leftrightarrow (1|0),\ Z\leftrightarrow (0|1)\ Y\leftrightarrow (1|1).
\end{equation}
An operator $ P= X \otimes Z $ then corresponds to $ (10|01) $.  

The symplectic inner product between $P\rightarrow(x|z)$ and $Q\rightarrow(x'|z')$ is defined as 
\begin{equation}
    \Omega(P,Q)=x \cdot z'+z\cdot x'(\text{mod} \ 2).
\end{equation}
The symplectic inner product determines whether two Pauli strings $ P, Q $ commute or anticommute.  The commutation and anticommutation criteria can then be expressed as 
\begin{align}
    &\Omega(P,Q)=0,\ \Longleftrightarrow [P,Q]=0.\\
    &\Omega(P,Q)=1,\ \Longleftrightarrow \{P,Q\}=0.
\end{align}

Using symplectic representation, the independence of a stabilizer set can be evaluated through its binary rank. For a stabilizer set in Pauli form $Q=\{P_1,\dots,P_m\}$, the corresponding binary matrix can be defined as 
\begin{equation}
    M_Q=
    \begin{pmatrix}
        \boldsymbol{x}_1|\boldsymbol{z}_1\\
        \boldsymbol{x}_2|\boldsymbol{z}_2\\
        \vdots\\
        \boldsymbol{x}_m|\boldsymbol{z}_m
    \end{pmatrix}.\label{matrixh}
\end{equation}
Here $\boldsymbol{x}_i, \boldsymbol{z}_i \in \mathbb{F}_2^N$ are binary row vectors encoding the $X$ and $Z$ components of $P_i$, respectively. The binary rank is 
\begin{equation}
    D_Q=\text{rank}_{\mathbb{F}_2}(M_Q),
\end{equation}
which gives the number of independent Pauli strings in $Q$ modulo global phases.

Gaussian elimination over $\mathbb{F}_2$ extracts a basis for the binary span of $Q$ using row swaps and row additions modulo $2$.
Columns are processed from left to right, selecting a pivot row and eliminating nonzero entries below the pivot through
\begin{equation}
    R_j\leftarrow R_j \oplus R_i\  (\text{mod} \ 2),\label{xor}
\end{equation} 
where $R_i$ is the pivot row and $\oplus$ denotes bitwise XOR. Columns without an available pivot are skipped.
After reduction to row-echelon form, the nonzero rows form a basis, and their number is $D_Q$~\cite{gottesman1997stabilizer, aaronson2004improved}.

To retain the original Pauli strings, we process them sequentially and keep $P_i$ only when its binary row $\textbf{v}_i=(x_i|z_i)$ increases the rank, i.e.,
\begin{equation}
    \text{rank}_{\mathbb{F}_2}    \begin{pmatrix}
        M_{\text{pre}}\\
        \textbf{v}_i
    \end{pmatrix}> \text{rank}_{\mathbb{F}_2}M_{\text{pre}},\label{rankbig}
\end{equation}
where $M_{\text{pre}}$ contains the previously retained original rows.
If $Q$ is mutually commuting, the selected subset remains mutually commuting.

The binary representation ignores global phases, and the row addition corresponds to Pauli multiplication modulo these phases. Multiplication phases must therefore be tracked separately when reconstructing signed operators \cite{dehaene2003clifford}. The signs of the selected generators are assigned in the sign assignment procedure discussed below.

\subsection{Stabilizer ground state}

The stabilizer ground states \cite{sun2025stabilizer} are defined as the stabilizer states which minimize the energy expectation value of a given Hamiltonian $H$:
\begin{align}
|S_{\text{SGS}}\rangle 
\in 
\operatorname*{arg\,min}_{|S \rangle \in {\cal S} }
\langle S | H | S \rangle,
\end{align}
where the minimization is performed over all $ N $-qubit stabilizer states $|S\rangle$ in the set $ \cal S $ with $ D_S $ elements.
The minimizer need not be unique.
The energy of the SGS state is given by 
\begin{align}
&E_{\text{SGS}} = \min_{|S \rangle \in {\cal S} }
E(|S\rangle,H),
\label{sgsenergy}
\end{align}
where $E(|S\rangle,H)=\langle S | H | S \rangle$.

It has been shown that for a 1D local Hamiltonian, the stabilizer ground state can be obtained with linear 
 computational complexity \cite{sun2025stabilizer}. Furthermore, the stabilizer ground state can be an efficient initial state for variational quantum algorithms (VQAs) \cite{leonetti2025deterministic}.

In general, the stabilizer ground state is not unique. 
Accordingly, we define the $N$-qubit stabilizer ground-state manifold as
\begin{equation}
\mathcal{S}_{\mathrm{SGS}}
=
\left\{
|S \rangle \in {\cal S} 
:
E(|S\rangle,H)
=E_{\text{SGS}} 
\right\},
\label{sgsset}
\end{equation}
where the minimization is taken over all $N$-qubit stabilizer states.  We define the degeneracy of the SGS manifold as 
\begin{align}
D_{\text{SGS}} = | \mathcal{S}_{\mathrm{SGS}} | .
\end{align}

\subsection{Optimal stabilizer ground state}
\label{sec:osgsdef}

As given in (\ref{sgsset}), the stabilizer ground state is potentially non-unique.  
Within $ \mathcal{S}_{\mathrm{SGS}} $, various stabilizers may have different fidelities with respect to the true ground state subspace. 
In many applications, the figure of merit of an initial state is high fidelity, rather than merely a low energy state.  For example, when using quantum phase estimation to find Hamiltonian eigenstates, one obtains a given eigenstate with the probability corresponding to its fidelity. We therefore define the {\it optimal stabilizer ground states} (OSGSs) as the stabilizer ground state that obtaining the highest fidelity with the true ground state subspace.

To give the formal definition of OSGS, let us first define the projector onto the ground state subspace of $H$ as 
\begin{equation}
\Pi_0=\sum_{f=0}^{d_0-1} |E_{0,f}\rangle\langle E_{0,f}|.
\end{equation}
Here, $d_0=\text{Tr}(\Pi_0) $ is the dimension of the ground state subspace.  The fidelity of between true ground state subspace and given stabilizer state can be defined as
\begin{equation}
    {\cal F}(|S\rangle,|E_0\rangle)=\langle S|\Pi_0|S\rangle.
\end{equation}

Then the set of optimal stabilizer ground states is defined as
\begin{align}
    &\mathcal{S}_{\text{OSGS}}=\operatorname*{arg\,max}_{|S_{\text{SGS}}\rangle\in\mathcal{S}_{\text{SGS}}}\langle S_{\text{SGS}}|\Pi_0|S_{\text{SGS}}\rangle.\label{stateosgs}
\end{align}
Thus where $\mathcal{S}_{\text{SGS}}$ contains the stabilizer state with the lowest stabilizer group energy,  $ \mathcal{S}_{\text{OSGS}} $  selects those with the highest ground state subspace fidelity.

If OSGS is degenerate, any state in $\mathcal{S}_{\text{OSGS}}$ can serve as an optimal stabilizer state candidate according to this criterion.
We note that the number of OSGS does not necessarily equal to $d_{\text{OSGS}}$.  Only {\it independent} OSGS group contributes to $d_{\text{OSGS}}$. 
In particular, mutually orthogonal OSGS groups belonging to distinct symmetry sectors may be directly associated with different degenerate sectors.

\subsection{Reformulation of generator selection problem}

The most direct way to find the SGS or OSGS is by running through all the stabilizer states, and checking each of their energy expectation values or ground state fidelities, respectively.    However, due to the superexponential number of stabilizer states, it is more desirable to find an alternative, more efficient method.  To this end, we turn to stabilizer methods.  
For a given Hamiltonian $H$, the exact OSGS preparation problem can be reformulated as the search for a signed optimal generating set $g_{\text{OSGS}}$ of the form given in Eq.~(\ref{generators}).  
Here we provide some general properties that $g_{\text{OSGS}}$ must satisfy.
 
To specify a unique pure stabilizer state, $g_{\text{OSGS}}$ must satisfy the following constraints.
Here, its elements are signed Pauli strings $s_iP_i$, with $s_i\in \{\pm1\}$. 

\begin{enumerate}[label=\roman*)]
\item Commutativity:
\begin{equation}
    [P_i,P_j]=0,\ \forall P_i, P_j\in g_{\text{OSGS}}.\label{commut}
\end{equation} \\  
\item Linear Independence:
\begin{equation}
    \text{rank}_{\mathbb{F}_2}(M_{g_{\text{OSGS}}})=|g_{\text{OSGS}}|,
\end{equation}
where the rows of $M_{g_{\text{OSGS}}}$ are the binary representations of the generators, with their overall signs omitted. \\
\item Completeness:
\begin{equation}
    |g_{\text{OSGS}}|=N.
\end{equation} \\
\item Two optimization objectives:
The OSGS stabilized by $g_{\text{OSGS}}$ minimizes the stabilizer energy defined in Eq.~(\ref{sgsenergy}). Among the states attaining this minimum, it maximizes the fidelity with the true ground state as satisfied in Eq.~(\ref{stateosgs}).
These objectives defien the exact OSGS target are applied sequentially.
\end{enumerate}

\section{Algorithm for selecting the Optimal Stabilizer Group Set}\label{sec3}

In this section, we reformulate the preparation problem of OSGS within the language of stabilizer formalism and introduce the allowed operations and definitions used in the following procedure. 
We note that in this section we only describe the procedure itself.  Its verification, along with worked examples, are described in Sec.~\ref{sec4}.

\subsection{Algorithm overview}

Here we give a brief summary of the algorithm that is used to generate the OSGS.  We first give an overview and further details will be given in subsequent sections.

\begin{enumerate}
    \item  For a given Pauli Hamiltonian, construct the Hamiltonian support set ${\cal P}_H$ according to Eq.~(\ref{ph}).
    
    \item  Construct the commutation matrix $C^H$ with Eq.~(\ref{commuth}) and map it to a vertex-weighted commutation graph, omitting self-loops.
    
    \item Use a genetic algorithm with the cost function Eq.~(\ref{fitmaxc}) to select a candidate mutually commuting subset pool $Q_{\text{pool}}$ which can be used for candidate generation.

    \item Extract an independent basis from ${\cal Q}^{(i)}$, solve the weighted parity constraint optimization problem using Eq.~(\ref{exmin}), and evaluate the resulting candidate energy. Iterate Steps 3-4 using the energy-based refinement Eq.~(\ref{fiteng}) to guide genetic selection and retain the lowest-energy candidates found.
    
     \item Apply the sufficiency criterion in Eq.~(\ref{degcri}) to identify symmetry-induced degeneracy of the Hamiltonian.
     
     \item Select compatible commuting centralizer subsets ${\Gamma}_{\text{can}}^{(\alpha)}$ using a genetic algorithm with objective function Eq.~(\ref{fitadd}). 
     
    \item Assign signs to $q_{\text{can}}$, augment it with independent centralizer generators while retaining $q_{\text{can}}$, and assign the remaining signs using the auxiliary-Hamiltonian and phase-assignment procedures in Eq.~(\ref{canass}).
    
    \item Prepare the selected state using a Clifford state preparation circuit constructed from its stabilizer tableau.
\end{enumerate}

Steps 1-2 construct the Hamiltonian support and its commutation graph.
Steps 3-4 form an iterative candidate search: Eq.~(\ref{fitmaxc}) provides the initial weight-based candidates, whereas subsequent genetic selection Eq.~(\ref{fiteng}) is guided by the energy obtained from sign assignment.
Steps 5-6 identify symmetry-induced degeneracy and select the compatible centralizer subsets.
Step 7 perform dependence reduction, parity-constraint reformulation, and sign assignment.
Step 8 completes the generating sets, and prepares the final candidate using the stabilizer tableau method.

We now give further details of each step.

\subsection{Step 1: Hamiltonian pre-processing}

Consider an $N$-qubit system. The Hamiltonian $H$ can be expressed as 
\begin{align}
    &H=\sum_i h_i P_i^H,\label{ham}
\end{align}
where $P_i^H\in {\cal P}$ denotes a Hermitian Pauli string without additional signs, and the real coefficient $h_i$ carries the sign of the corresponding Hamiltonian term.   If the Hamiltonian is not given in this form, the coefficients can be computed by 
\begin{equation}
    h_i=\frac{1}{2^N}\mathrm{Tr}(P_i^H H)\label{bi},
\end{equation}
which follows from the orthogonality relation $\mathrm{Tr}(P_i P_j ) = 2^N \delta_{ij}$.

The non-zero coefficient Pauli strings which are not the identity serve as candidates for stabilizer generators. Let us call these Pauli strings as the Hamiltonian support
\begin{align}
{\cal P}_H = \{P_i^H:h_i\neq 0, P_i^H\neq I^{\otimes N} \},\label{ph}
\end{align}
where $D_H=|{\cal P}_H|$.  The identity $I^{\otimes N}$ belongs to every full  stabilizer generator group but is not an independent generator.
It is excluded from ${\cal P}_H$ because $h_0 I$ only produces a constant energy shift, where $h_0 $ is the coefficient of the identity $ I^{\otimes N} $.   

The exact stabilizer energy is determined by these terms, their coefficients, and their signs in the full stabilizer group $G$, together with the constant contribution $h_0$.
Membership must therefore be checked in $G$, rather than the generating set $g$.

\subsection{Step 2: Commutation-graph construction}

To construct a candidate generating set for $g_{\text{OSGS}}$  based on the Hamiltonian support ${\cal P}_H$, we define a commutation matrix $C^H$ and use it to construct a vertex-weighted undirected commutation graph $C$.  

The commutation matrix $C^H$ characterizes the pairwise commutation relations between the Pauli strings in ${\cal P}_H$:  
\begin{equation}
    C_{ij}^H=
    \left\{
\begin{array}{ll}
     1, & \text{if }\   [P_i^H,P_j^H]=0, \\[4pt]
     0, & \text{if }\   [P_i^H,P_j^H]\neq 0.
\end{array}
\right.\label{commuth}
\end{equation}
Since $C_{ii}^H=1$, we omit the diagonal entries when constructing the graph.

We define the commutation graph $C$ with the information in $C^H$. 
Each vertex of $C$ represents a Pauli string in ${\cal P}_H$, and an edge connects two distinct vertices if the corresponding Pauli strings commute.  
The weight of each vertex is given by the magnitude of the corresponding Hamiltonian coefficient:
\begin{equation}
    w_{i}=|h_i|,\ P_i^H \in {\cal P}_H.\label{weights}
\end{equation}
Thus the signs of $h_i$ are not encoded in the commutation graph $C$.  
An example of a commutation graph $C$ for the transverse-field Ising model is shown in Fig.~ \ref{fig:maxcli}.

\subsection{Step 3: Weighted commuting-subset search} 

We employ a genetic algorithm on the commutation graph $C$ to generate a candidate commuting subsets pool.  
The proxy objective in Eq.~(\ref{fitmaxc}) is used for the initial weight-based search. Subsequent genetic refinement is guided by the sign optimized energies evaluated in Step 4.

In the genetic algorithm, each individual represents a candidate subset of Pauli strings.
We encode each individual as a binary chromosome
\begin{equation}
    \textbf{x}=\{x_1,\dots,x_{D_{H}}\},\ x_i\in\{0,1\}.
\end{equation}
Here, $x_i=1$ indicates that the Pauli string $P_i^{H}$ is selected, whereas $x_i=0$ indicates that it is not selected. 
The chromosome length is $D_H=|{\cal P}_H|$.
The set of Pauli matrices themselves that corresponds to the selected subset is therefore 
\begin{equation}
    g(\textbf{x})=\{P_i^H\in{\cal P}_H : x_i=1\}.
\end{equation}
This set is a candidate subset, not necessarily an independent generating set.
The encoding specifies only whether each Pauli string is selected; the associated signs are determined later, in the sign assignment steps.

We define the proxy cost function to be minimized as
\begin{align}
    F_{\text{proxy}} = & -M\sum_{i=1}^{D_H}|h_i|x_i-\beta \sum_{i=1}^{D_H} x_i +\eta\sum_{i<j}(1-C_{ij}^H)x_ix_j .\label{fitmaxc} 
\end{align}
Here, $h_i$ is the signed coefficient of $P_i^H$ in the Hamiltonian as defined in Eq.~(\ref{bi}).
The first term favors subsets with larger total weights, while the second term, with $\beta>0$, favors larger subsets when their total weights are comparable. 
The positive parameter $M$ and $\beta$ control the relative importance of total weights and subset size, with $M$ chosen to emphasize the former.

The last term in (\ref{fitmaxc}) is a penalty term that penalizes infeasible individuals containing non-commuting Pauli strings. 
In particular, if $x_i=x_j=1, \ C_{ij}^H=0$, the pair contributes a positive penalty $\eta$.
The parameter $\eta$ is chosen sufficiently large that a single commutation violation outweighs the maximum combined reward from total wights and subset size. Specifically, we choose
\begin{equation}
    \eta> M\sum_{i=1}^{D_H} |h_i|+\beta D_H.
\end{equation}
This condition ensures that every infeasible individual has a higher cost that any feasible individual.

The genetic algorithm starts from a randomly generated population, and the cost function is evaluated for each individual.
Individuals with lower $F_{\text{proxy}}$ are preferentially selected, 
while crossover and mutation generating new generating subsets. Before energy evaluation, candidates are repaired to be mutually commuting and augmented to inclusion-maximal subsets of ${\cal P}_H$. 

{Instead of retaining only the best individual, we retain multiple distinct feasible candidates with low proxy costs to preserve structural diversity. 
The best feasible candidate found under the proxy objective is retained as a seed, while other candidate structures remain available for subsequent refinement. The resulting candidate pool is defined as
\begin{equation}
    Q_{\text{pool}}=\{{\cal Q}^{(1)},{\cal Q}^{(2)},\dots,{\cal Q}^{(K)}\},\ {\cal Q}^{(a)}=g(\mathbf{x}^{(a)}).
\end{equation}
Here $K$ is the number of retained distinct candidates.

For a candidate subset ${\cal Q}^{(i)}$, we define
\begin{equation}
    H_{{\cal Q}i}=\sum_{\mu=1}^m h_\mu P_{\mu},\ m=|{\cal Q}^{(i)}|.
\end{equation}
each ${\cal Q}^{(i)}$ is a low weight-based candidate rather than certified stabilizer ground state solutions. They may contain linearly dependent elements or contain less then $N$ components which will effect the exact energy evaluation.
Hence, Step 4 extract an independent basis from candidates in $ Q_{\text{pool}}$ which has the lowest energy.

\subsection{Step 4: Energy refinement}
In this step, we formulate the energy minimization over the generator signs as a weighted parity-constraint optimization problem. The attained energies guide further genetic refinement and determine which candidate branches are retained for subsequent stabilizer completion. 

\subsubsection{Dependence reduction of ${\cal Q}^{(i)}$ }
 
With Eq.~(\ref{matrixh}), each candidate ${\cal Q}^{(i)}$ is transferred into its binary matrix $M_{{\cal Q}i}$. We use Eq.~(\ref{xor}) to eliminate the dependent Pauli terms from ${\cal Q}^{(i)}$.  
The retained independent unsigned basis is 
\begin{equation}
    q({\cal Q}^{(i)})=\{B_1,\dots,B_r\}.
\end{equation}
The corresponding binary matrix is represented as $M_{qi}$.

\subsubsection{Reconstruction of $H_{{\cal Q}i}$ } 

Except for the sign, each Pauli term in $H_{{\cal Q}i}$ can be expressed as a product of elements in $q({\cal Q}^{(i)})$ which can be expressed as 
\begin{equation}
     P_{\mu}=\eta_{\mu} \prod_{j=1}^{r}B_j^{A^s_{\mu j}},\eta_{\mu}\in\{\pm1\}.\label{pmu}
\end{equation}
$\eta_{\mu}$ is determined by separately tracking the multiplication sign, following the stabilizer formalism and sign-aware Pauli representation \cite{gottesman1997stabilizer,dehaene2003clifford}.
$A^s_{\mu j}\in \mathbb{F}_2$ indicates whether the basis $B_j$ appears in the product which can be obtained by solving 
\begin{equation}
    A^s M_{qi}=M_{{\cal Q}i}.
\end{equation}
This procedure constructs an $m\times r$ binary matrix $A^s$ that specifies the generator combination representing each Pauli term up to an overall sign.

For a subset sign configuration $z\in(z_1,z_2,\dots,z_r),\ z\in \mathbb{F}_2^r$, the sign is determined by
\begin{equation}
    s_j=(-1)^{z_j}. \label{sjxj}
\end{equation} 
The corresponding energy contribution can be defined as \cite{kirby2020classical}
\begin{align}
    &E_{{\cal Q}i}(z)=E_I+\sum_{\mu=1}^m J_\mu (-1)^{A^s_{\mu}z},\label{energform}\\
    &A^s_{\mu}z=\sum_{j=1}^r A^s_{\mu j}z_j\pmod 2,\label{amux}
\end{align}
where $J_\mu=h_\mu \eta_\mu$.
$E_I$ denotes the constant energy shift in $H$. 
Since $E_I$ does not affect the minimizing sign configuration, it is omitted during optimization.

\subsubsection{Reformulation of the sign assignment optimization problem } 

For a fixed candidate ${\cal Q}^{(i)}$, Eq.~(\ref{energform}) gives the energy associated with the sign configuration $z$. 
In this formula, the sign of each term's energy contribution depends jointly on $A^s_{\mu}z$ and its sign $\mu_j$.
For each Pauli term $P_\mu$, we define the preferred parity as
\begin{equation}
    b_\mu=
     \left\{
    \begin{array}{ll}
     1, & \   J_\mu>0, \\[4pt]
     0 &\ J_\mu<0.
\end{array}
\right.
\end{equation}
Violating this preference increases the energy contribution by $2|J_\mu|$.
Therefore, minimizing $E_{{\cal Q}i}(z)$ is equivalent to minimizing the total weight of violated parity constraints. 

Equation (\ref{energform}) can be reformulated as 
\begin{align}
    &E_{{\cal Q}i}(z)=E_I-\sum_{\mu=1}^m w_\mu +2L_{{\cal Q}i}(z),\label{exmin}\\
    &L_{{\cal Q}i}(z)=\sum_{\mu=1}^m w_\mu [ 1- \delta(A^s_\mu z=  b_\mu)]
\end{align}
Here, $w_\mu=|J_\mu|$, and the quantity in the square brackets works as a indicator function to return the parity violation weight. It is equal to $1$ when the parity constraint is violated and $0$ otherwise.
The sign assignment problem is then formulated as $ \min_{z\in\mathbb{F}_2}L_{{\cal Q}i}(z)$.

\subsubsection{Genetic optimization of signs}
For a fixed candidate ${\cal Q}^{(i)}$, the sign configuration is encoded by a binary chromosome $z$.
The inner search minimizes
\begin{equation}
    F_{\text{sign}}(z|{\cal Q}^{(i)})=L_{{\cal Q}i}(z).\label{fitsign}
\end{equation}
Starting from a randomly generated population, configurations with lower cost are preferentially selected. Crossover and bit-flip mutation generate new configurations, while elitism preserves the best configuration found. The lowest cost sign configuration found is retained together with its associated candidate. In general, a finite inner search does not certify the optimal sign configuration. 

Let ${\cal Z}_{\text{search}}(Q^{(i)})$ denote the sign configurations evaluated for ${\cal Q}^{(i)}$. The energies-based cost function for the outer search is
\begin{equation}
    F_{\text{energy}}({\cal Q}^{(i)})=\min_{z\in {\cal Z}_{\text{search}}(Q^{(i)})} E_{{\cal Q}i}(z). \label{fiteng}
\end{equation}
}

Steps 3-4 form an outer iterative genetic loop. The proxy objective in Eq.~(\ref{fitmaxc}) initializes the candidate population. Step 4 evaluates each candidate through the inner sign search. 
Whenever a lower energy is found, the candidate, its sign configuration, and its energy are recorded as the best result so far.
Candidates with lower $F_{\text{energy}}({\cal Q}^{(i)})$ are preferentially selected as parents, while the best result is preserved by elitism. 
Crossover and mutation of the support selection chromosomes in Step 3 generate new candidate subsets. These operations modify the selected generating subset. Then the independent basis and parity constraints are reconstructed. Sign crossover is performed only within a fixed candidate.

The outer loop terminates when the generation budget is exhausted or the best energy remains unchanged for a prescribed number of generations. Let ${\cal G}_{\text{Search}}$ contain all candidate subsets evaluated during the search.
Upon the termination, the lowest energy candidate found is retained as
\begin{equation}
    {\cal Q}_{\text{can}}\in \arg \min_{{\cal Q}^{(i)}\in {\cal G}_{\text{Search}}}F_{\text{energy}}({\cal Q}^{(i)}).
\end{equation}
The associated sign configuration is retained for sunbsequent stabilizer completion.

The Hamiltonian component associated with ${\cal Q}_{\text{can}}$ is
\begin{equation}
H_{\text{com}}=\sum_{P_i^H\in {{\cal Q}_{\text{can}}}} h_i^{\text{com}} P_{i}^H\label{hcom}.
\end{equation}
Here, $|{\cal Q}_{\text{can}}|=m$, so $H_{\text{com}}$ contains $m$ Pauli terms.
We note that the lowest eigenvalue of $H_{\text{com}}$ corresponds to $E_{\text{SGS}}$, as will be explained below. 

By subtracting $H_{\text{com}}$ from $H$, we define the auxiliary Hamiltonian as
\begin{equation}
H_\text{aux} =H-H_{\text{com}},
\label{haux}
\end{equation}
Its Pauli coefficients are denoted by $h_i^{\text{aux}}$, and its nonidentity support is denoted by ${\cal P}_{H_{\text{aux}}}$, such that
\begin{align}
H_\text{aux}=\sum_{P_i^H \in {\cal P}_{H_{\text{aux}}} } h_i^{\text{aux}}  P_{i}^H \label{hauxexpansion}.
\end{align}
%

\subsection{Step 5: Symmetry-induced degeneracy criterion}

The candidate subset $\cal Q_{\text{can}}$ is in general insufficient to uniquely specify the OSGS.  Additional 
generators may be required, as well as their signs, which depends in part on the degeneracy of $H$.
In this subsection, we identify the Pauli centralizers elements and establish a sufficient criterion for symmetry-induced energy-level degeneracy.

The phase-free Pauli centralizer of $H$ is the set of linearly independent Pauli strings in $\cal P$ that commute with $H$, defined as
\begin{equation}
    \Gamma(H)=\{P\in {\cal P}:\ [P,H]=0\}.\label{cengroup}
\end{equation}
Modulo global phases, generators of $\Gamma(H)$ can be obtained from the symplectic orthogonal complement of the binary span of ${\cal{P}}_H$ by Gaussian elimination over $\mathbb{F}_2$\cite{gottesman1997stabilizer,calderbank1998quantum}.

The centralizer can be found by solving for the null space by Gaussian elimination.   Using the binary symplectic representation introduced in Sec. \ref{subsla}, a Pauli string $P$ belongs to $\Gamma^H$ if and only if
\begin{align}
    & M_H\Lambda\begin{pmatrix}
        \textbf{x}^T\\
        \textbf{z}^T
    \end{pmatrix} =0,\label{vch}
    \end{align}
where
\begin{align}
    &\Lambda=
    \begin{pmatrix}
        0&I_N\\I_N&0
    \end{pmatrix}.
\end{align}
$M_H$ represents the binary matrix of ${\cal P}_H$. 
The solution of Eq.~(\ref{vch}) may contains linearly dependent terms, therefore Gaussian elimination is implemented to reduce the dependency.

A sufficient condition for every energy eigenspace of $H$ to be at least twofold degenerate is
\begin{align}
    \exists P,Q \in \Gamma(H), \{P,Q\}=0.\label{degcri}
\end{align}
This criterion is proven in Appendix~\ref{appena}.

\subsection{Step 6: Centralizer augmentation}

As shown in Appendix ~\ref{appenb}, projection onto appropriately chosen centralizer eigenspaces can preserve the ground state fidelity after normalization, making compatible centralizer elements suitable candidates for augmenting ${\cal Q}_{\text{can}}$ toward a complete generating set $g_{\text{OSGS}}$.
In this subsection, we seek a maximum-cardinality mutually commuting subset of the compatible centralizer pool.

The centralizer pool compatible with $\mathcal{Q}_{\text{can}}$ is defined as 
\begin{equation}
    \Gamma(H|\mathcal{Q}_{\text{can}})=\{P\in \Gamma(H):[P,Q]=0,\forall Q \in \mathcal{Q}_{\text{can}}\}.\label{chgcan}
\end{equation}
Although all the elements in $\Gamma(H|\mathcal{Q}_{\text{can}})$ commute with every element of $\mathcal{Q}_{\text{can}}$, they are not guaranteed to commute with one another.
We therefore construct the commutation graph $C^{{\Gamma}}$ 
and apply a genetic algorithm to search for mutually commuting subsets of maximum cardinality. 
The commutation matrix $C^{{\Gamma}}$ is constructed using the same rule as Eq.~(\ref{commuth}) for pairs of Pauli strings in the centralizer candidate set.
The corresponding graph $C^{\Gamma}$ is constructed in the same way, but with all vertex weights set to $1$.

The cost function to be maximized is
\begin{align}
     F_{\max}&=M\sum_{i=1}^{D_{\Gamma}} x_i-\eta\sum_{i<j} (1-C_{ij}^{\Gamma})x_ix_j .
     \label{fitadd} 
\end{align}
Here, $D_{\Gamma}=|\Gamma(H|\mathcal{Q}_{\text{can}})|$, and $x_i\in\{0,1\}$ indicates whether the $i$th Pauli string in the pool is selected.
We choose $ M>0, \eta>M D_{\Gamma}$, so that every infeasible individual has a lower objective value than any feasible individual.
The second term in (\ref{fitadd}) is a penalty function to remove non-commuting pairs in the same way as (\ref{fitmaxc}).

The distinct feasible candidate subsets retained from the search are collected as
\begin{align}
&\Gamma_{\text{can}}=\{\Gamma_{\text{can}}^{(1)},\dots,\Gamma_{\text{can}}^{(D_{\text{can}})}\}, 
\end{align}
where $D_{\text{can}}$ is the number of retained subsets.
Different commuting centralizer subsets ${\Gamma}_{\text{can}}^{(\alpha)}$, with $\alpha=1,\dots,D_{\text{can}}$, define different candidate stabilizer branches.
Each subset is combined with $\mathcal{Q}_{\text{can}}$, and the combined set is subsequently reduced to independent generators in the following subsection.

\subsection{Step 7: Sign assignment}

To obtain the final signed generating set $g_{\text{OSGS}}$, we progressively remove linear independencies and assign consistent signs to the generators in $q_{\text{can}}$ and the augmented candidate set $\widetilde{q}_{\text{can}}^{(\alpha)}$. The $g_{\text{OSGS}}$ is obtained by removing linear dependencies from $\widetilde{q}_{\text{can}}^{(\alpha)}$ and assigning consistent signs to the retained generators.
%
%
The procedure of sign assignment is performed in three steps as described below.

\subsubsection{Sign assignment of $q_{\text{can}}$.}\label{secsign}

According to Eq.~(\ref{exmin}), minimizing $E(z)$ is equivalent to minimizing the total weight of violated parity constraints.
The general weighted parity-constraint optimization problem is a NP-hard problem \cite{johan2001some,berlekamp1978inherent}. We retain its sign configuration in Step 4.

\subsubsection{Dependency reduction of $\widetilde{q}_{\text{can}}^{(\alpha)}$.}

If $q_{\text{can}}$ contains $r<N$ independent generators, compatible centralizer elements provide candidates for augmentation.
We first combine the set of $q_{\text{can}}$ and $\Gamma_{\text{can}}^{(\alpha)}$. In the binary matrix Eq.~(\ref{matrixh}) construction procedure, elements in $q_{\text{can}}$ will be put in the first rows.
Then we implement Gaussian elimination Eq.~(\ref{xor}) to it. 
The retained group is defined as $\widetilde{q}_{\text{can}}^{(\alpha)}$ with dimension $D_{\text{can}}^{(\alpha)}=|\widetilde{q}_{\text{can}}^{(\alpha)}|$.
This set can be a good candidate for OSGS group.

\subsubsection{Sign assignment of $\widetilde{q}_{\text{can}}^{(\alpha)}$.}

The signs of the retained centralizer generators are assigned using $H_{\text{aux}}$.
Applying the genetic algorithm Eq.~(\ref{fitmaxc}) to search for the maximal commuting subset of $H_{\text{aux}}$, we obtain a candidate subset ${\cal Q}_{\text{can}}^{\text{aux}}$ and implement Gaussian elimination Eq.~(\ref{xor}) to obtain the subset $q_{\text{can}}^{\text{aux}}$. 

The signs $t_j$ of the auxiliary generators are determined by solving the corresponding parity-constraint optimization Eq.~(\ref{exmin}). Some specific solutions are discussed in step 1 in Sec. \ref{secsign}.  

For each retained, unassigned centralizer $P_i^{(\alpha)}$ that admits the presentation
\begin{equation}
    P_i^{(\alpha)} =\tau_i \prod_{j} P_{\text{aux}}^{(j)},\ \tau_i\in \{\pm1\}, \ P_{\text{aux}}^{(j)}\in q_{\text{can}}^{\text{aux}},
\end{equation}
we assign its sign as 
\begin{equation}
    s_i=\tau_i\prod_j t_j.\label{canass}
\end{equation}
$\tau_i$ accounts for the sign arising from the product of Pauli strings. 
Any remaining unassigned signs of the retained generators are chosen randomly to complete the candidate sign configuration.

\subsection{Step 8: OSGS preparation}
If the signed generating set has rank $D_{\text{can}}^{(\alpha)}<N$, we add $N-D_{\text{can}}^{(\alpha)}$ generators such that the complete set is mutually commuting and binary-independent.
Either generator satisfies Eq.~(\ref{commut}) can be added into such generating subset to make it complete.
This can be done by computing equation similar to Eq.~(\ref{vch}), where $M$ is a binary matrix of the existing generators, and applying symplectic Gram-Schmidt reduction to obtain logical Pauli pairs. 
Selecting one operator from each pair provides the required additional generators \cite{wilde2009logical}.
The signs of these additional generators can be chosen randomly, with each choice specifying a unique pure stabilizer state within the subspace defined by the existing signed generators.

Once the signed OSGS generating set is obtained, a Clifford circuit for preparing the corresponding state can be constructed using stabilizer tableau representation \cite{de2011linearized, aaronson2004improved}. 
By encoding the generators of the initial state $|0\rangle^{\otimes N}$ and the target OSGS in tableau form, and applying tableau transformations corresponding to Clifford gates, including the associated sign updates, one can construct a circuit that maps the initial state to the OSGS. 
The tableau formalism also supports efficient classical simulation of stabilizer circuits \cite{gidney2021stim, masot2024stabilizer}.


\section{Analysis and illustration of algorithmic steps}
\label{sec4}

In this section, we analyze the properties of the algorithm introduced in the previous section, and verify it for the simple example of the transverse-field Ising model.

\subsection{Step 1: Hamiltonian support}
Suppose the pure state $|S_g\rangle$ has the full signed stabilizer group $G$. 
The energy expectation value for the Pauli-decomposed Hamiltonian is  
\begin{equation}
   E(|S_g\rangle,H)=\sum_{i} h_i \langle S_g| P_i^H |S_g\rangle.\label{lstaben}
\end{equation}
For a pure stabilizer state, each Pauli expectation value satisfies 
\begin{align}
    \langle S_g|P_i^H |S_g\rangle=\left\{ 
\begin{array}{ll}
     1, & \   P_i^H\in G, \\[4pt]
     -1, &\ -P_i^H\in G,
     \\[4pt]
     0, & \text{otherwise}.
\end{array}
\right.\label{expvalsps}
\end{align}
A related group-theoretic characterization of stabilizer state energy was also employed by Robin~\cite{robin2025stabilizer} in selecting an optimal stabilizer Hamiltonian using an energy-minimization criterion.

Equation (\ref{expvalsps}) shows that, apart from the identity contribution, only Hamiltonian-support terms that belong to $G$ up to a sign contribute to the stabilizer energy. 
The first step therefore constructs the Hamiltonian support ${\cal P}_H$, which contains all nonidentity Pauli strings with nonzero coefficients in $H$, as the initial pool of candidate generators. Membership in ${\cal P}_H$ does not imply a nonzero expectation value in a given stabilizer state.
The generator signs are not assigned at this stage, they are optimized jointly in the sign assignment procedure performed later.

\paragraph*{Example:} To illustrate the algorithm, we consider the four-qubit transverse-field Ising model with open boundary conditions 
\begin{equation}
    H=\sum_i^{N-1} Z_iZ_{i+1}+\lambda\sum_j^N X_j .  \label{case}
\end{equation}
For any $ \lambda  \ne 0 $, the Hamiltonian support is
\begin{equation}
    {\cal P}_H=\{Z_1Z_2,Z_2Z_3,Z_3Z_4,X_1,X_2,X_3,X_4\}.\label{phcase}
\end{equation}
When $\lambda=0$, the Hamiltonian support is 
\begin{equation}
    {\cal P}_H=\{Z_1Z_2,Z_2Z_3,Z_3Z_4\}.\label{phcase0}
\end{equation}

\subsection{Step 2: Commutation graph}

In order to find $g_{\text{OSGS}}$, we must identify mutually commuting subsets of \(\mathcal P_H\) with large total coefficient magnitudes. 
In Eq.~(\ref{lstaben}), the stabilizer state energy is expressed as a weighted sum of the Pauli expectation values, with coefficient $h_i$. 
The stabilizer generators must commute with one another, as expressed in Eq.~(\ref{commut}). 

These observations motivate a maximum-weight commuting subset search as an energy proxy optimization. 
In this framework, mutually commuting subsets correspond to the clique problem in the associated undirected graph \cite{wu2015review}. 
To address this problem, we construct the commutation matrix $C^H$ for ${\cal P}_H$, the vertex weights assigned according to Eq.~(\ref{weights}).
The cardinality reward in Eq.~(\ref{fitmaxc}) additionally favors larger subsets.
The resulting candidates require subsequent sign optimization and full energy evaluation.

\paragraph*{Example:} Continuing with the transverse field Ising model in Eq.~(\ref{case}), the corresponding commutation matrix $C^H$ is 
\begin{table}[t]
\caption{Commutation matrix $C^H$ for Hamiltonian in Eq.~(\ref{case}). Entries are $1$ for commuting pairs of Pauli strings and $0$ for anticommuting pairs.}
\label{tabcommut}
\centering
\begin{ruledtabular}
\begin{tabular}{c*{7}{c}}
 & $Z_1Z_2$ & $Z_2Z_3$ & $Z_3Z_4$ & $X_1$ & $X_2$ & $X_3$ & $X_4$ \\
 \hline
  $Z_1Z_2$ & 1 & 1 & 1 & 1 & 0 & 0 & 1 \\
  $Z_2Z_3$ & 1 & 1 & 1 & 1 & 0 & 0 & 1 \\
  $Z_3Z_4$ & 1 & 1 & 1 & 1 & 1 & 0 & 0 \\
  $X_1$     & 0 & 1 & 1 & 1 & 1 & 1 & 1 \\
  $X_2$     & 0 & 0 & 1 & 1 & 1 & 1 & 1 \\
  $X_3$     & 1 & 0 & 0 & 1 & 1 & 1 & 1 \\
  $X_4$     & 1 & 1 & 0 & 1 & 1 & 1 & 1
\end{tabular}
\end{ruledtabular}
\end{table}

Since $C^H$ is symmetric, we use only its strictly upper triangular entries to construct the graph.
The vertex weights follow from Eq.~(\ref{bi}).

\begin{figure}[t]
    \centering
    \includegraphics[width=1.0\linewidth]{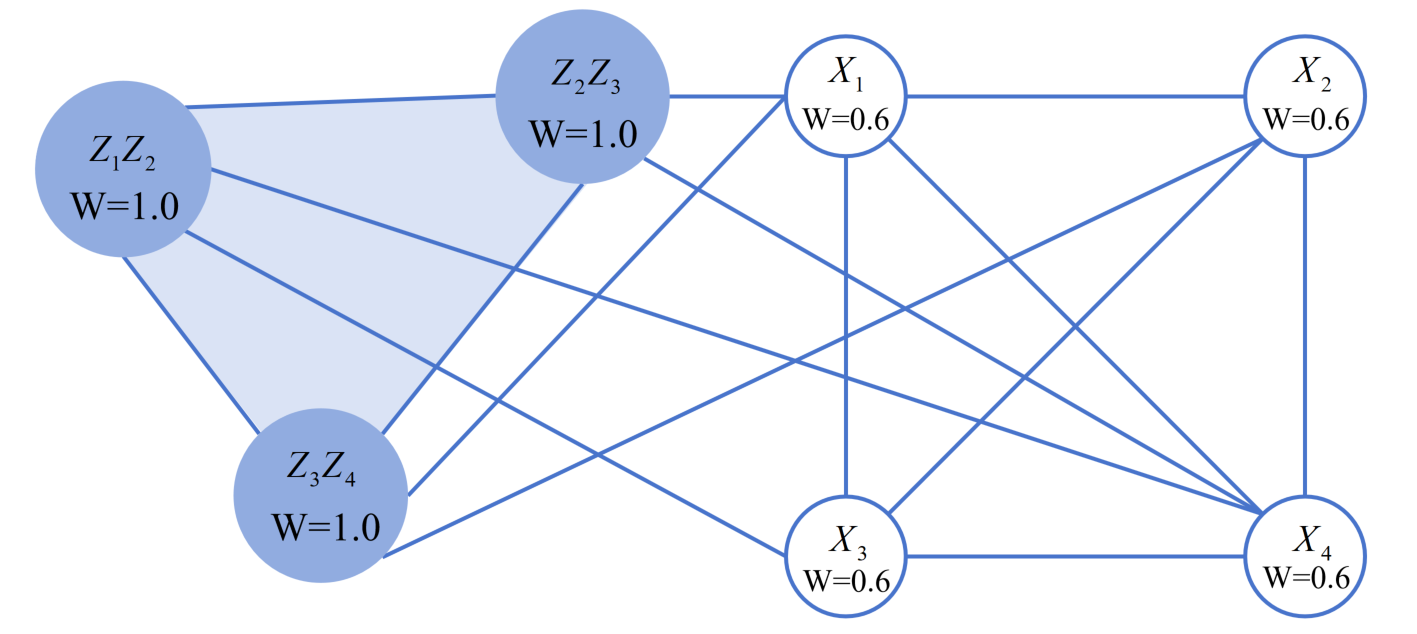}
    \caption{Commutation graph $C$ for the $4$-qubit system with Hamiltonian (\ref{case}).  We take the parameter $ \lambda = 0.6 $. }
    \label{fig:maxcli}
\end{figure}

In Fig.~\ref{fig:maxcli}, we show the weighted maximal clique formulation of the candidate commuting subset search.
For this example, the maximum weight clique is $\{Z_1Z_2,Z_2Z_3,Z_1Z_3\}$, with total weight 3.

The commutation graph encodes the pairwise commutation relations and coefficient magnitudes in a single representation.
However, a clique does not necessarily form a binary independent or complete generating set, and the graph does not encode generator signs.
Dependence reduction, sign optimization, and necessary generating set completion are considered in the next steps.

\subsection{Step 3: Maximal commuting stabilizer subset search}

Finding a maximum weight clique is an NP-hard problem.
Polynomial time approximation algorithms are available for the unweighted maximum-clique problem, with guarantees that depend on graph size \cite{feige2004approximating}.
Various classical algorithms have been developed for clique search problems, including greedy algorithms \cite{butenko2003maximum}, local search algorithms \cite{wu2013adaptive}, and neutral network-based methods \cite{cheng2011finding}. 
By reformulating the maximum-weight clique problem into a quadratic unconstrained binary optimization problem and mapping it to an Ising model, quantum algorithms such as quantum annealing and the Quantum Approximate Optimization Algorithm (QAOA) can be applied to these formulations \cite{chapuis2017finding, herrman2021impact}.
Evolutionary algorithms, such as genetic algorithms, have also been applied to the maximum weighted clique problem \cite{guturu2008impatient}. 

We employ a genetic algorithm to search for the highest weight mutually commuting subsets, avoiding explicit enumeration of all candidate subsets.
Through iterative selection, crossover, and mutation over successive generations, the algorithm searches for candidates with lower values of the energy proxy cost function in Eq.~(\ref{fitmaxc}), which combines weight and cardinality rewards with penalties for commutation violations.
It returns a candidate pool $Q_{\text{pool}}$ which will further selected in Step 4.

\paragraph*{Example:} In the case of Eq.~(\ref{case}), $D_H=7$. A binary chromosome is defined by Eq.~(\ref{phcase}). The corresponding candidate pool is 
\begin{equation}
    Q_{\text{pool}}=\{\{Z_1Z_2,Z_2Z_3,Z_3Z_4\},\{X_1,X_2,X_3,X_4\}\}.
\end{equation}
Here there are two candidate found, the corresponding sign assignment and energy-based refinement will be given in Step 4.

\subsection{Step 4: Weighted parity constraint problem}

We now explain how the sign assignment problem of candidate can be reformulated to an energy minimization problem of its Hamiltonian terms.

During dependence reduction, we obtain the linearly independent generator group $q({\cal Q}^{(i)})$.
The Pauli terms in ${\cal Q}^{(i)} \setminus q({\cal Q}^{(i)})$ can be expressed as the product of elements in $q({\cal Q}^{(i)})$, up to multiplicative signs $\eta_\mu$, as in Eq.~(\ref{pmu}).
Based on Eq.~(\ref{expvalsps}), their eigenvalues are therefore determined by the signs assigned to the independent generators and cannot be chosen independently.

Using these product relations and the simultaneous eigenstate condition, the corresponding energy is written as Eq.~(\ref{energform}). 
The parity of Eq.~(\ref{amux}) determines whether the factor $(-1)^{A_\mu x}$ has a positive or negative value.
The effective coefficient $J_{\mu}=h_{\mu} \eta_{\mu}$ incorporates both the Hamiltonian coefficient and the Pauli multiplication sign.
Its sign determines the preferred parity $b_{\mu}$.
When this preference is satisfied, the term contributes $-|J_\mu|$, otherwise, it contributes $+|J_\mu|$, increasing the energy by $2|J_\mu|$.
As expressed in Eq.~(\ref{exmin}), minimizing the total weight of violated parity constraints is therefore equivalent to minimizing $E(z)$ for the fixed candidate subset.

For a fixed candidate ${\cal Q}^{(i)}$, Eq.~(\ref{exmin}) shows that minimizing its energy is equivalent to minimizing the weighted parity constraint loss. The sign configuration is encoded by a binary chromosome $z$, and the inner genetic search uses the cost in Eq.~(\ref{fitsign}) to identify the best sign configuration found within search budget.
Eq.~(\ref{fiteng}) then gives the attained energy for this candidate.
Comparing these energies over all evaluated candidates selects the lowest energy subset, denoted by ${\cal Q}_{\text{can}}$, together with its associated sign configuration.

In Eq.~(\ref{hcom}), we collect the Hamiltonian terms associated with ${\cal Q}_{\text{can}}$. 
For a state $|\psi_x\rangle$ in the simultaneous eigenspace specified by the generator sign configuration $x$, the sign assignment objective for this component is
\begin{equation}
    E_{\text{com}}^{\min}=\min_{x\in \mathbb{F}_2^r} \langle \psi_x|H_{\text{com}}|\psi_x\rangle.
\end{equation}
From Eq.~(\ref{energform}) we have $\min_x E(z)=E_{\text{com}}^{\min}$ in which the identity's contribution is omitted. 
This is equal to the minimization full Hamiltonian expectation value $E_{\text{SGS}}$ as all nonidentity terms have zero expectation value in Eq.~(\ref{expvalsps}).

\paragraph*{Example:} For the transverse field Ising model Eq.~(\ref{case}), the candidate pool has two candidate generating subsets.
For candidate ${\cal Q}^{(1)}=\{Z_1Z_2,Z_2Z_3,Z_3Z_4\}$, $m=r=3$, and $A^s$ is expressed as 
\begin{equation}
    A^s=\begin{pmatrix}
1 & 0 & 0 \\
0 & 1 & 0 \\
0 & 0 & 1
\end{pmatrix}.
\end{equation}
For all $\mu$, $J_\mu=1$ and the preferred parity is $b_\mu=1$.
The task of minimizing Eq.~(\ref{exmin}) can be expressed as finding a valid sign configuration $z=(z_1,z_2,z_3)$ that minimizes $L(z)$. Eq.~(\ref{fitsign}) gives the result as $z=(1,1,1)$.

For ${\cal Q}^{(2)}=\{X_1,X_2,X_3,X_4\}$, $m=r=4$, all $\mu,J_{\mu}=1$ such that $b_\mu=1$ as well. The sign configuration that minimizing Eq.~(\ref{fitsign}) is given as $z=(1,1,1,1)$.

For $\lambda=0$, ${\cal Q}_{\text{can}}=\{Z_1Z_2,Z_2Z_3,Z_3Z_4\}$. $H_{\text{com}}=Z_1Z_2+Z_2Z_3+Z_3Z_4$, $H_{\text{aux}}=0$. In this case, $E_{\text{com}}^{\min}=E_{\text{SGS}}=-3$. 

For any  $0<|\lambda|<0.75 $, we have ${\cal Q}_{\text{can}}=\{Z_1Z_2,Z_2Z_3,Z_3Z_4\}$. The corresponding common and auxiliary Hamiltonian are then
\begin{align}
    &H_{\text{com}}=Z_1Z_2+Z_2Z_3+Z_3Z_4,\\
    &H_{\text{aux}}=\lambda (X_1+X_2+X_3+X_4). \label{hauxcase}
\end{align}
$E_{\text{com}}^{\min}=E_{\text{SGS}}=-3$ equals to the SGS energy.

For the case of $ |\lambda| \geq 0.75 $, ${\cal Q}_{\text{can}}=\{X_1,X_2,X_3,X_4\}$ with $E_{\text{com}}^{\min}=E_{\text{SGS}}=-4\lambda$. 
Its corresponding common and auxiliary Hamiltonian are 
\begin{align}
    &H_{\text{com}}=\lambda (X_1+X_2+X_3+X_4),\\
    &H_{\text{aux}}=Z_1Z_2+Z_2Z_3+Z_3Z_4.
\end{align}

\subsection{Step 5: Symmetry-induced degeneracy}
The signs assigned to centralizer generators determine the selected symmetry sectors and their compatibility with the ground state.

\paragraph*{Example:} For the Hamiltonian Eq.~(\ref{case}), the degeneracy changes with the value of $\lambda$.
We consider three different cases, $\lambda = 0 $, $ 0<|\lambda| < 0.75$ and $\lambda \geq 0.75 $.

For $\lambda=0$, $\Gamma(H)$ contains $31$ elements. We exclude $I^{\otimes N}$ elements. In this set, we can find independent symplectic conjugate centralizer pairs as $\{X_1 X_2 X_3 X_4,Z_i\}_{i=1}^4$ which indicates the Hamiltonian is degenerate. 
The different choice of $Z_i$ does not effect the number of conjugate pairs in $\Gamma(H)$.

For $ 0<|\lambda| < 0.75$, the centralizer group is $\Gamma(H)= \{X_1 X_2 X_3 X_4\}$. There is no conjugate centralizer pair in $\Gamma(H)$ which indicate the ground state is nondegenerate.

For $ |\lambda| \geq 0.75$, the centralizer group is again $\Gamma(H)= \{X_1 X_2 X_3 X_4\}$, which indicates the ground state is nondegenerate.

\subsection{Step 6: Augmented centralizer}
The candidate ${\cal Q}_{\text{can}}$ may contain fewer than $N$ independent generators and therefore may  be insufficient to uniquely specify a pure stabilizer state $|S_{\text{OSGS}}\rangle$.
The centralizer provides a pool of augmentation candidates.
The motivation follows from the fidelity argument presented in Appendix~\ref{appenb}.
For $P\in{\Gamma}(H)$, its eigenprojectors preserve each eigenspace of $H$. 
If the target ground state is an eigenstate of $P$, projection on the matching eigenvalue sector followed by normalization does not decrease the ground state fidelity, provided that the initial overlap is nonzero.

We therefore construct ${\Gamma}(H|{\cal Q}_{\text{can}})$ by selecting the elements of ${\Gamma}(H)$ that commute with every element of ${\cal Q}_{\text{can}}$. 
This does not guarantee that elements in ${\Gamma}(H|{\cal Q}_{\text{can}})$ commute with each other.
The task is to search for a maximum cardinality mutually commuting subset of ${\Gamma}(H|{\cal Q}_{\text{can}})$, using a procedure similar to the search for ${\cal Q}_{\text{can}}$.
The objective function in Eq.~(\ref{fitadd}) rewards larger subsets and penalizes commutation violations, while the genetic algorithm provides candidate solutions without guaranteeing global optimality.

The selected subset may not be unique. Different subsets ${\Gamma}_{\text{can}}^{(\alpha)}$ define different candidate augmentation branches.
Each branch is subsequently combined with ${\cal Q}_{\text{can}}$ and reduced to independent generators.

\paragraph*{Example:}  For the degenerate Hamiltonian case with $\lambda=0$ in Eq.~(\ref{case}), $\Gamma(H|{\cal Q}_{\text{can}})$ contains $15$ elements. $\Gamma_{\text{can}}$ can be written as
\begin{align}
    \Gamma_{\text{can}}^{(1)}=\{&Z_1Z_2,Z_2Z_3,Z_3Z_4,Z_1Z_4,Z_1Z_3,Z_2Z_4,Z_1Z_2Z_3Z_4,\nonumber\\
    &X_1X_2X_3X_4,X_1X_2Y_1Y_2,X_1Y_2X_3Y_4,X_1Y_2Y_3X_4,\nonumber\\
    &Y_1X_2X_3Y_4,Y_1X_2Y_3X_4,Y_1Y_2X_3X_4,Y_1Y_2Y_3Y_4\}.\label{gamacan0}
\end{align}

In the nondegenerate case with $\lambda\neq 0$, $\Gamma(H|{\cal Q}_{\text{can}})=\Gamma_{\text{can}}^{(1)}=\{X_1 X_2 X_3 X_4\}$.

\subsection{Step 7: Sign assignment}
The sign assignment is performed on independent generators, while the signs of dependent Pauli strings follow from their multiplication relations.

\subsubsection{Sign assignment of $q_{\text{can}}$}

The generator signs in $q_{\text{can}}$ are determined by the weighted parity-constraint optimization as described in Eq.~(\ref{exmin}).

\paragraph*{Example:} In the case of Eq.~(\ref{case}) with $|\lambda|<0.75$. 
By choosing the sign configuration as $z=(1,1,1)$, all the parity preferences will be satisfied and the corresponding sign is $s_i=-1$. 
The following signed generator set is $q_{\text{can}}=\{-Z_1Z_2,-Z_2Z_3,-Z_3Z_4\}$.

For the case of $|\lambda|\geq 0.75$, all the parity preference can be satisfied if we choose $z=(-1,-1,-1,-1)$. The corresponding $q_{\text{can}}$ is 
\begin{equation}
    q_{\text{can}}=\{-X_1,-X_2,-X_3,-X_4\}. \label{qcanbig}
\end{equation}
This equals to $g_{\text{OSGS}}$.

\subsubsection{Dependence reduction of $\widetilde{q}_{\text{can}}^{(\alpha)}$ }

Some centralizer elements may be linearly dependent on the generators in $q_{\text{can}}$ in the binary representation. These elements are removed during the dependence reduction.
This dependence reduction facilitates the subsequent auxiliary-based sign assignment while preserving the binary span of the candidate set.

\paragraph*{Example:} For the degenerate case $\lambda=0$ in Eq.~(\ref{case}), we have 
\begin{equation}
    \widetilde{q}_{\text{can}}^{(1)}=\{-Z_1Z_2,-Z_2Z_3,-Z_3Z_4, s_\Gamma P_\Gamma\},\  \forall P_{\Gamma}\in \Gamma_{\text{can}}^{(1)},\label{wildqcase}
\end{equation}
where $\Gamma_{\text{can}}^{(1)}$ is given in Eq.~(\ref{gamacan0}).
In this degenerate case, each element in Eq.~(\ref{gamacan0}) satisfies the requirements of commutativity.
And the element number coincides with the rank result in Eq.~(\ref{rankbig}) as $D_q=4$.

For the nondegenerate case with $0<|\lambda|<0.75$, the Gaussian elimination is implemented into the combined subset of $q_{\text{can}} \cup \Gamma_{\text{can}}$, and the elements $\{Z_1, Z_2,Z_3,Z_4\}$ can be produced by the multiplication of elements in $q_{\text{can}}$ such that they will be removed.
Then $\widetilde{q}_{\text{can}}^{(1)}$ takes the same form as Eq.~(\ref{wildqcase}). 

For the nondegenerate case with $|\lambda|\geq0.75$, $q_{\text{can}}$ in Eq.~(\ref{qcanbig}) satisfies $D_q=4$ which already satisfied the completeness of $g_{\text{OSGS}}$. So the Gaussian elimination of $q_{\text{can}} \cup \Gamma_{\text{can}}$ will not increase the number of elements 

\subsubsection{Sign assignment of $\widetilde{q}_{\text{can}}^{(\alpha)}$ }

For the retained additional generators that can be expressed using $q_{\text{can}}^{\text{aux}}$, signs are assigned according to Eq.~(\ref{canass}).
This provides an auxiliary based selection rule which maximizes the ground state fidelity. 
The resulting configurations are therefore treated as candidates for subsequent evaluation.

\paragraph*{Example:} For the degenerate case $\lambda=0$ in Eq.~(\ref{case}), $\widetilde{q}_{\text{can}}^{(1)}$ can be written as Eq.~(\ref{wildqcase}). The signs of the elements in Eq.~(\ref{qcanbig}) can be chosen randomly as it distinguishes different symmetry sectors within the ground state subspace. The change of sign will not effect the fidelity or energy.

For the nondegenerate case with $0<|\lambda|<0.75$, $H_{\text{aux}}$ is given by Eq.~(\ref{hauxcase}), and ${\cal Q}_{\text{can}}^{\text{aux}}=q_{\text{can}}^{\text{aux}}=\{X_1,X_2,X_3,X_4\}$ which is linearly independent. 
By solving the weighted parity constraint optimization for $H_{\text{aux}}$, the sign configuration is $z^{\text{aux}}=(1,1,1,1)$.
The centralizer can be produced by $X_1X_2X_3X_4=\prod_iX_i$ where $\tau_i=1$.
With Eq.~(\ref{canass}), the sign of the compatible centralizer is determined. 
And the resulting subset can be expressed as $g_{\text{OSGS}}=\{-Z_1Z_2,-Z_2Z_3,-Z_3Z_4, X_1X_2X_3X_4\}$.

For the case of $|\lambda| \geq 0.75$, $g_{\text{OSGS}}$ is Eq.~(\ref{qcanbig}).

\section{Application to quantum simulation}\label{sec5}

In this section, we demonstrate the utility of our approach to find the OSGS by using it as the initial state of measurement-based imaginary time evolution (MITE).  MITE allows for the preparation of an unknown ground state of a given Hamiltonian.  By using the OSGS as the initial state, we show that the convergence time can be improved considerably.

\subsection{The measurement-based imaginary time evolution method}

MITE combines weak energy measurements with conditional unitary operations to prepare low-energy states \cite{mao2023measurement}.
Let us briefly review the mechanism relevant to OSGS initialization.

In standard imaginary time evolution (ITE), the excited state components are suppressed with evolution time.  Provided that the initial state has nonzero overlap with a nondegenerate ground state, we have
\begin{equation}
     \lim_{\tau \rightarrow \infty} \frac{e^{-H\tau}|\psi_{\text{in}}\rangle}{||e^{-H\tau}|\psi_{\text{in}}\rangle||} \rightarrow |E_0\rangle.
\end{equation}
For a nondegenerate ground state, this limit equals to $|E_0\rangle$ up to a global phase.
For the case where there is a ground state degeneracy, the ITE will asymptotically drive the initial state into the ground state subspace. 
The initial state can be written as the superposition of energy eigenstates
\begin{align}
|\psi_{\text{in}}\rangle=\sum_nA_n|E_n\rangle,
\end{align}
where $\sum_n |A_n|^2=1$. 

In the MITE algorithm, a random sequence of weak measurements is combined with conditional unitary operations. Their measurement operators are 
\begin{align}
    M_0&=\sum_n \frac{1}{\sqrt{2}}\left(\cos{\epsilon E_n}-\sin{\epsilon E_n}\right)|E_n \rangle \langle E_n|, \\ 
    M_1&=\sum_n \frac{1}{\sqrt{2}}\left(\cos{\epsilon E_n}+\sin{\epsilon E_n}\right)|E_n \rangle \langle E_n|,
\end{align}
which satisfy $M_0^{\dagger}M_0+M_1^{\dagger}M_1=I$. 
For $||\epsilon H||\ll 1, \ \epsilon>0$, they approximate $e^{-\epsilon H}/2$ and $e^{+\epsilon H}/2$, respectively.
These can be implemented with a Suzuki-Trotter decomposition when the required interaction is not directly available \cite{suzuki1993improved,trotter1959product,kapit2012non}.

Weak measurements modify the amplitude distribution of the input state. 
For a measurement sequence without feedback corrections, the measurement-induced filtering envelope becomes approximately Gaussian near its peak after numerous measurements.
The resulting state amplitudes also depend on the initial coefficients $A_n$.
Thus, the random measurements filter the state in the energy eigenbasis.  

The accumulated measurement outcomes allow us to estimate the approximate energy selected by the measurement sequence. 
If this estimate reaches or exceeds the energy threshold $E_{\text{th}}$, a unitary correction $U_C$ is applied to rotate the current state, and the measurement counters are reset.
With a threshold that separates the ground state energy from the excited spectrum, together with suitable unitary corrections, this feedback is used to steer the evolution toward the ground state.
We note that for a degenerate ground space, the procedure does not distinguish states in energy, and in general the final state depends upon the particular measurement sequence and initial state. 

The number of measurements required depends on the Hamiltonian structure, initial state configuration and the target accuracy.
Repeated measurement sequences and feedback operations contribute to the quantum resource cost.
Each weak-measurement operation uses an ancillary qubit, which can be reset and reused in subsequent steps.
An initial state with a small overlap with the ground state may require more measurement and correction steps. 
A suitable initial state can therefore reduce the cost of convergence of the MITE algorithm.
We investigate the OSGS as such an initial state and evaluate its effect on MITE.

\subsection{OSGS as an initial state for MITE}

The goal of obtaining the OSGS is to accelerate the MITE scheme by preparing an optimized initial state using the stabilizer formalism.
The candidate state is selected offline using classical search and sign assignment procedures described in the previous sections.
It is then prepared using a Clifford circuit and supplied as the initial state for MITE.

The motivation can be understood from its energy distribution.
The optimal stabilizer ground state attains the lowest stabilizer-group energy
\begin{align}
   E_{\text{SGS}}=\langle \psi_{\text{OSGS}}|H|\psi_{\text{OSGS}}\rangle .
\end{align}
Expanding the state in the energy eigenbasis gives
\begin{align}
|\psi_{\text{OSGS}}\rangle & =\sum_n A_n |E_n\rangle \\ 
     E_{\text{SGS}}& =\sum_n |A_n|^2E_n,
\end{align}
where $A_n=\langle \psi_{\text{OSGS}}|E_n\rangle$ denotes the state overlap.

The effect of initialization can be examined through the weak measurement filtering process.
For a sequence containing $k_0$ outcomes associated with $M_0$ and $k_1$ outcomes associated with $M_1$, with no conditional unitary corrections, the normalized output is
\begin{equation}
    |\psi_k\rangle=\frac{M_0^{k_0}M_1^{k_1}|\psi_{\text{OSGS}}\rangle}{||M_0^{k_0}M_1^{k_1}|\psi_{\text{OSGS}}\rangle||}=\sum_n\widetilde{A}_n^{(k)}|E_n\rangle,\label{psik}
\end{equation}
provided that the denominator is nonzero, where $k=k_0+k_1$ and $\epsilon E_n\in [-\pi/4,\pi/4]$.
The normalized energy resolved amplitude is defined as 
\begin{equation}
    \widetilde{A}_n^{(k)}=\frac{A_nf_K(\epsilon E_n)}{\sqrt{\sum_n|A_n|^2|f_K(\epsilon E_n)|^2}},\label{amplitude}
\end{equation}
which satisfies $\sum_n|\widetilde{A}_{m}^{(k)}|^2=1$.  Here we defined
\begin{equation}
    f_K(\epsilon E_n)=\cos^{k_0}(\epsilon E_n+\frac{\pi}{4})\sin^{k_1}(\epsilon E_n+\frac{\pi}{4}).
\end{equation}
The ground state weight after this sequence is 
\begin{equation}
|\widetilde{A}_{0}^{(k)}|^2=\frac{|A_0|^2|f_K(\epsilon E_0)|^2}{\sum_n|A_n|^2|f_K(\epsilon E_n)|^2}.
\end{equation}
This expression shows how the initial energy distribution combines with the measurement filter.
A favorable initial ground state overlap can reduce the filtering required, although the overall convergence also depends on measurement outcomes.
Once the estimate energy exceeds the threshold, 
the unitary correction will reset the state into OSGS and restart the MITE iteration.  

Finally, the OSGS-initialized MITE can be used for ground state energy estimation.  Once the output state is sufficiently close to the ground state subspace, its energy can be estimated by measuring the Hamiltonian expectation value.

\subsection{Complexity analysis and convergence error}

We now estimate the complexity of the OSGS-based MITE procedure.  The quantum resource cost consists of two parts: (i) preparation of selected OSGS, and (ii) the subsequent MITE scheme.
We discuss the gate counts and sequential measurement rounds; the classical preprocessing cost is not included in these estimates.

Once the signed generating set is known, it can be stored using $N(2N+1)$ bits, and its Clifford preparation circuit can be constructed using the tableau methods described above.
Any unitary stabilizer circuit has an equivalent implementation with $O(N^2/\log N)$ elementary Clifford gates \cite{aaronson2004improved,mohseni2022ising}.
This is a gate count bound, the circuit depth additionally depends on gate scheduling and hardware connectivity.

Under the specified restart protocol, the expected total cost required to prepare a state with ground state fidelity at least $1-\delta$ is bounded by
\begin{equation}
    \mathbb{E}[T_{\delta}]\leq \frac{C_{\text{prep}}+k_\delta}{f_0 a_0^{k_\delta}} ,
\end{equation}
where
\begin{align}
  f_0& =\mathcal{F}(|E_0\rangle,|\psi_{\text{OSGS}}\rangle)  \\
  a_0& =\cos^2(\epsilon E_0+\frac{\pi}{4}) .
\end{align}
Here, the cost of each weak measurement round is normalized to one, and $C_{\text{prep}}$ denotes the cost of OSGS preparation at the start of each attempt, expressed in the same units. 
$k_\delta$ is the sufficient number of consecutive $M_0$ outcomes defined in Eq.~(\ref{meatimesande0}). 
Further details are given in Appendix~\ref{appenc}.

For the ideal convergence analysis, we assume a nondegenerate target ground state and define the convergence error
\begin{equation}
    \mathcal{E}_K=1-\mathcal{F}(|E_0\rangle,|\psi_K\rangle),
\end{equation}
where $|\psi_K\rangle$ is normalized.
$\mathcal{E}_K$ measures convergence error, excluding hardware noise and implementation errors.

Consider a continuous sequence of $K$ weak measurements without intervening unitary corrections, starting from a normalized state $|\psi_{\text{OSGS}}\rangle$ with nonzero ground state overlap.
Under the weak measurement filtering approximation, with the amplitude peak near $E_{\text{th}}$, 
The upper bound of convergence error after $k$ iterations is given as
\begin{equation}
 \mathcal{E}_K\geq \frac{\mathcal{F}(|\psi_{\text{OSGS}}\rangle,|E_1\rangle)}{\mathcal{F}(|\psi_{\text{OSGS}}\rangle,|E_0\rangle)}e^{2K\epsilon^2(E_0-E_1)(E_0+E_1-2E_{\text{th}})}. \label{conerror}
\end{equation} 

Equation (\ref{conerror}) is an asymptotic estimate for this filtering regime which is determined by the excited-to-ground fidelity ratio, the number of weak measurements, and spectral properties of $H$.
When OSGS initialization reduces this ratio, it lowers the error prefactor and can reduce the measurement count needed to reach a given accuracy under the same filtering conditions.
Further details and the assumption underlying the approximation are given in Appendix~\ref{append}.

\section{Validation and numerical simulation}
\label{sec:val}

\subsection{Validation for small systems}

For systems with $N\leq 5$ , exhaustive enumeration of the pure stabilizer state using the matrix representation employed in robustness of magic calculations provides an exact benchmark for SGS selection. 
This approach is resource intensive because the number of pure stabilizer states grows rapidly with $N$.
We use this enumeration to validate our OSGS search algorithm for small $ N $. 

The robustness of magic (RoM) \cite{PhysRevLett.118.090501} quantifies the extent to which a quantum state lies outside the convex hull of pure stabilizer states.
It can be formulated as 
\begin{equation}
    R(\rho)=\min || \vec{x} ||_1\  \text{subject to}\  A_{\text{stab}}\vec{x}  = \bm{b},
    \label{romlinear}
\end{equation}
where $|| \vec{x} ||_1=\sum_j  |x_j|$.  
For a quantum state $\rho$, elements of $\bm{b}$ are $b_i=\mathrm{Tr}\left(P_i \rho\right)$.
The matrix $A_{\text{stab}}$ contains the Pauli expectation values of all pure stabilizer states with different Pauli signs.
This $ 4^N \times D_S $ matrix $ A $ is defined by
\begin{align}
[A_{\text{stab}}]_{ij}= \langle S_j |s_i P_i | S_j \rangle, 
\label{amatdef}
\end{align}
where $ \{|S_j\rangle\}_{j=1}^{D_S}$ is a set of pure $N$-qubit stabilizer states, and $\{s_iP_i\}_{i=1}^{4^N}$ is the Pauli basis, including the identity. Such $A_{\text{stab}}$ matrix up to $5$ qubits is given by the RoM method \cite{PhysRevLett.118.090501}.

Each column of $ A_{\text{stab}} $ corresponds to one stabilizer state. 
Its nonzero entries identify the elements of the associated full signed stabilizer group:
\begin{equation}
    \widetilde{G}_j=\{ [A_{\text{stab}}]_{ij}P_i:[A_{\text{stab}}]_{ij}\neq 0\},
\end{equation}
The minimization defining RoM is performed over
\begin{align}
    \vec{x} = (x_1, x_2, \dots, x_{D_S})^T .
    \label{xvecdef}
\end{align}
These real coefficients describe a stabilizer state decomposition (not to be confused with the binary selection variables used in the genetic algorithm).

For Hamiltonian (\ref{ham}), define
\begin{equation}
    b_i^H=\text{Tr}(P_iH).
\end{equation}
Then the SGS energy can be written as
\begin{align}
    E_{\text{SGS}}     & = \min_{j\in [1,D_S] } \frac{1}{2^N} \sum_{i=1}^{4^N} [A_{\text{stab}}]_{ij} b_i^H.
    \label{emins}
\end{align}
Equivalently, the vector of stabilizer state energies is 
\begin{align}
    \textbf{e}& =\frac{1}{2^N}A_{\text{stab}}^{T}\textbf{b}^H \\ 
    E_{\text{SGS}}& =\min_{j\in [1,D_S]} e_j.
\end{align}

Thus, the SGS energy minimization uses the stabilizer state matrix but does not require solving the RoM optimization problem.
Physically, $E_{\text{SGS}}  $ is the lowest attainable energy among all the pure stabilizer states.
If several states attain this energy, the OSGS is selected by maximizing their fidelity with the target ground state.

While the above is a perfectly consistent procedure, the size of the $A_{\text{stab}}$ limits the scalability of this approach. 
Explicit dense storage requires $O(4^ND_S)$ entries.
The original RoM implementation considered systems up to $N=5$ qubits \cite{PhysRevLett.118.090501}.
More advanced methods \cite{hamaguchi2024handbook} have enabled RoM calculations for arbitrary states up to $N=7$ and stabilizer fidelity calculations up to $N=8$.
Nevertheless, explicit enumeration of the complete matrix remains impractical for large systems.

\subsection{Example: The transverse-field Ising model}

To verify the OSGS selection algorithm, we consider an $N$ qubit transverse-field Ising model (\ref{case}) which serves as a natural platform for studying quantum phase transitions \cite{heyl2013dynamical}.  In the following discussion we set $N=4$ and restrict attention to $\lambda\geq0$.
The optimal stabilizer generating set depends on $\lambda$, and we examine the following cases.

Here we show the result obtained by the exhaustive search.
For $\lambda=0$, $E_{\text{SGS}}=-3$ is evaluated by Eq.~(\ref{emins}) and the OSGS $|\psi_{\text{OSGS}}\rangle=\{|0101\rangle,|1010\rangle,(\pm|0101\rangle+ |1010\rangle )/\sqrt{2},(\pm i |0101\rangle+|1010\rangle )/\sqrt{2}\}$ which is $6$ fold degenerate and all the fidelity is ${\cal F}(|\psi_{\text{OSGS}}\rangle,|E_0\rangle)=1$. 

For $0< |\lambda| < 0.75$, $E_{\text{SGS}}=-3$. The OSGS obtained by the exhaustive search is $|\psi_{\text{OSGS}}\rangle=(|0101\rangle+ |1010\rangle )/\sqrt{2}$.

For $|\lambda| > 0.75$, $E_{\text{SGS}}=-4\lambda$. The OSGS obtained by the exhaustive search is $|\psi_{\text{OSGS}}\rangle=|----\rangle$.

For the crossover point of $\lambda=0.75$, the $E_{\text{SGS}}=-3$ while the OSGS is $|\psi_{\text{OSGS}}\rangle=|----\rangle$.

\subsection{Numerical simulation}\label{sec7}

\begin{figure}[htbp]
    \centering
    \includegraphics[width=1.0\linewidth]{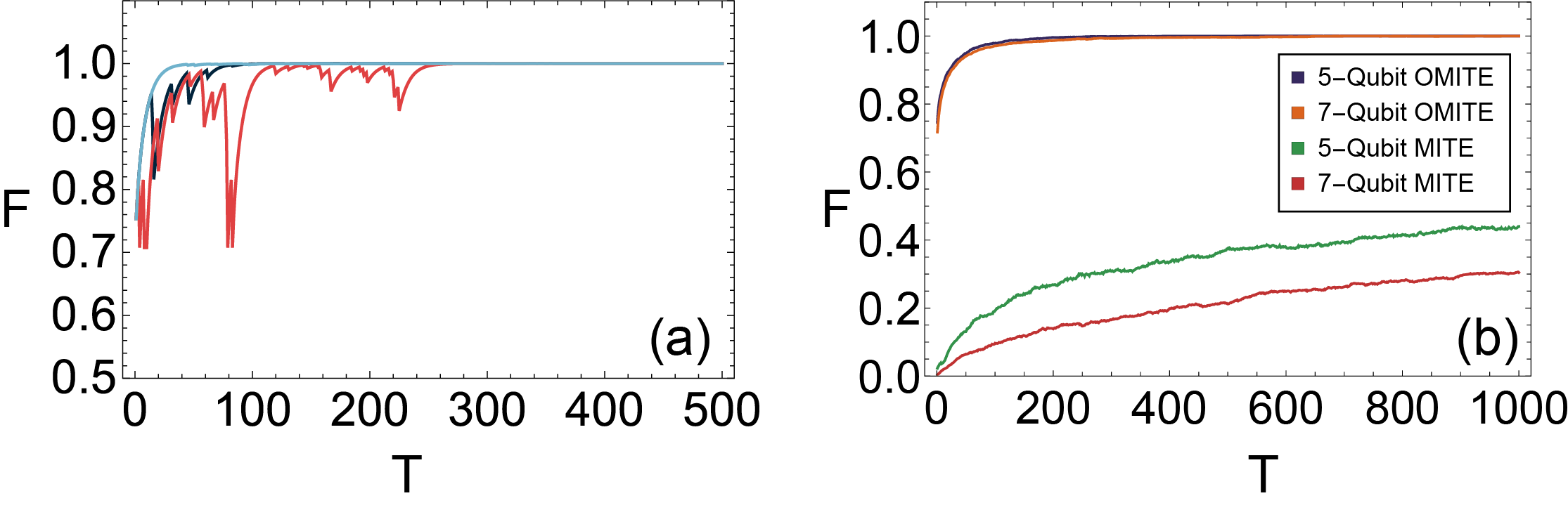}
    \caption{Performance of the OSGS-based MITE scheme applied to the transverse-field Ising model at $\lambda=0.6$.  
    (a) Solid curves in three colors show representative fidelity trajectories for the $N=5$ qubit system.
    (b) Comparison of the mean fidelity at each measurement step, averaged over 1000 independent trails in $N=5$ and $N=7$ systems. OMITE and MITE denote the ground state preparation with and without OSGS initialization, respectively. }
    \label{fig:simul}
\end{figure}
\begin{figure}[htbp]
    \centering
    \includegraphics[width=1.0\linewidth]{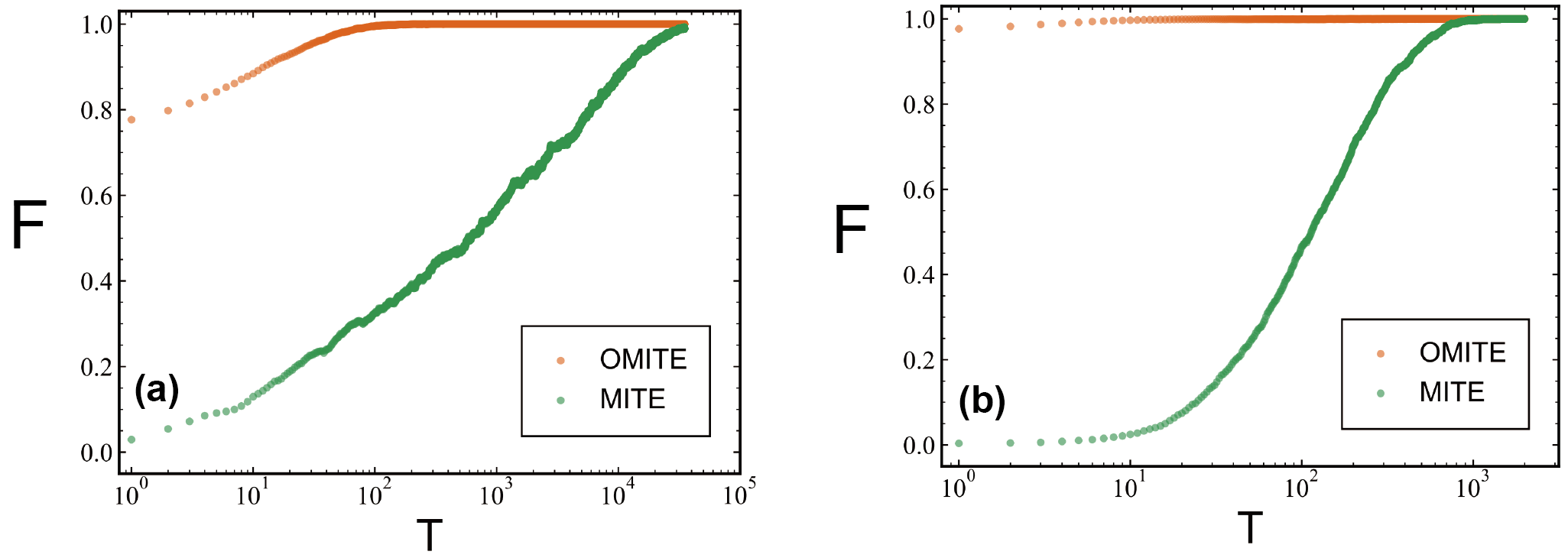}
    \caption{Comparison of MITE performance for 5-qubit transverse-field Ising model at (a) $\lambda=0.6$; (b) $\lambda=2 $. OMITE and MITE denote the ground state preparation with and without OSGS initialization, respectively.}
    \label{figdiflambda}
\end{figure}

We evaluate the performance of OSGS-based MITE for transverse-field Ising model.
The numerical results for representative measurement trajectories for the $N=5$ system at $\lambda=0.6$ are shown in Fig.~\ref{fig:simul}(a). 
Fig~\ref{fig:simul}(b) compares MITE with and without OSGS initialization for $5$-qubit and $7$-qubit systems. 
Using the OSGS as the initial state leads to a substantially faster convergence of the mean ground state fidelity in these simulations.
With the OSGS initialization, the mean fidelity covergence curves remain similar when the system size increases from $5$ to $7$ qubits.
By comparison, MITE without OSGS initialization exhibits slower convergence for the larger system under the simulated conditions.
As shown in Fig.~\ref{figdiflambda}, the improvement is also observed at both parameter values considered $\lambda=0.6$ and $\lambda=2$.

\section{Conclusions}\label{sec6}

In this work, we proposed a method for preparing the OSGS of an arbitrary Hamiltonian,  defined as the stabilizer ground state that has the maximum fidelity with the ground state. 
Finding the OSGS by exhaustive search is highly numerically intensive, due to the superexponential scaling of the number of stabilizer states with the number of qubits.  To address this issue, we introduced a concrete algorithm to find the OSGS using the stabilizer formalism.  Our algorithm searches for OSGS candidates for Hamiltonians expressed as sums of Pauli strings.  
Our method selects candidate commuting subsets from the Hamiltonian support using a genetic algorithm guided by an energy proxy objective.
These subsets are augmented with compatible centralizer elements, reduced to independent generators, and assigns consistent signs using the weighted parity constraint formulation and auxiliary Hamiltonian information. 
The centralizer-based criterion certifies symmetry-induced degeneracy and provides a lower bound on the eigenspace dimension.
After completing the generating set, the corresponding candidate stabilizer state is prepared using a Clifford circuit constructed from its stabilizer tableau. The method is a heuristic method and as such does not guarantee global optimality.  However, we have found that our algorithm is able to correctly identify the OSGS in all cases that we have numerically tested.

We demonstrate the application of the OSGS algorithm by using its output as an initial state for measurement-based imaginary time evolution (MITE).
Numerical simulation results show that OSGS initialization can significantly accelerate convergence toward the true ground state in the transverse-field Ising model that we considered. 
The improvement is expected since the OSGS serves as a high-fidelity initial state that is a good approximation to the true ground state.  Due to the high density of stabilizer states in Hilbert space, there is always a stabilizer state that is nearby that can serve as a high quality initial state. The complexity and error analyses clarify how the initial ground state overlap and filtering parameters affect the measurement cost and conditional convergence error.

This procedure provides a framework for stabilizer-assisted ground-state preparation with potential applications in quantum computation and quantum information. The method is particularly promising for quantum simulation applications where finding the many-body ground state is a generic problem. 
The primary advantage is that, once a candidate stabilizer state has been selected, it can be efficiently prepared using a Clifford circuit and can reduce the number of subsequent MITE measurements.
The candidate-state construction does not require prior knowledge of the exact energy spectrum, although identifying the highest fidelity candidate generally requires additional information about the target ground state.

While we have used the stabilizer formalism in constructing our OSGS algorithm, it is not restricted to Hamiltonians with stabilizer ground states.
The Clifford circuit prepares a stabilizer approximation to the ground state, which is subsequently refined by MITE.
The quality of this approximation affects the initial fidelity and the subsequent convergence rate, reliable ground state preparation therefore depends on both the initialization and MITE procedures.
Related methods use stabilizer decompositions and quasiprobability representations to quantify or simulate non-Clifford resources \cite{PhysRevLett.118.090501, bennink2017unbiased}.

\section*{ACKNOWLEDGMENTS}\label{sec9}

This research is supported by the Natural Science Foundation of Zhejiang Province under Grant No. LQN26A050002.
T.B. is supported by Tamkeen under the NYU Abu Dhabi Research Institute grant CG008.

\section*{DATA AVAILABILITY}

No data were created or analyzed in this study.


\begin{appendices}
\appendix
\renewcommand{\appendixname}{APPENDIX}     

\section{\MakeUppercase{Proof of degeneracy criterion}}\label{appena}
Anticommuting Pauli symmeties provides a sufficient condition for energy degeneracy. 
Specifically, if the Pauli centralizer $\Gamma(H)$ contains two Hermitian Pauli strings $P$ and $Q$ satisfying $\{P,Q\}=0$, then {\it every} energy eigenspace of $H$ has dimension of at least two.

Let $E$ be any eigenvalue of $H$. Since $[P,H]=0$ and $P$ is Hermitian, we can choose a normalized simultaneous eigenstate $|\psi\rangle$ satisfying
\begin{align}
    &H|\psi\rangle=E|\psi\rangle,\\
    &P|\psi\rangle=p|\psi\rangle,
\end{align}
where $p=\pm1$.

The definition of the centralizer ensures $[Q,H]=0$ such that
\begin{equation}
    H(Q|\psi\rangle)=QH|\psi\rangle=E(Q|\psi\rangle).
\end{equation}
Thus, $Q|\psi\rangle$ belongs to the same energy eigenspace. 
Moreover, $Q$ is unitary, and $Q|\psi\rangle$ remains normalized and is nonzero.
Using $\{P,Q\}=0$, we obtain 
\begin{equation}
    P(Q|\psi\rangle)=-Q(P|\psi\rangle)=-p\ Q|\psi\rangle.
\end{equation}
The state $|\psi\rangle$ and $Q|\psi\rangle$ have opposite eigenvalues with respect to the Hermitian operator $P$ and are orthogonal.
Since both have energy $E$, the eigenspace associated with $E$ has dimension at least $2$.

\section{\MakeUppercase{Effect of centralizer constraint on stabilizer state fidelity} }\label{appenb}

The fidelity between the normalized states $|S_{\text{SGS}}\rangle$ and $|E_0\rangle$ is
\begin{align}
    \mathcal{F}(|E_0\rangle, |S_{\text{SGS}}\rangle)
    &=\langle E_0 | S_{\text{SGS}}\rangle \langle S_{\text{SGS}} |E_0\rangle \nonumber \\
    &=\left\langle E_0 \middle| \prod_{j=1}^{N} \frac{I+s_jP_j^{(i)}}{2} \middle| E_0 \right\rangle. \label{fids}
\end{align}
Here, we the sign $s_j$ is chosen accordingly, $P_j$ is the mutually commuting, binary-independent generators of $|S_{\text{SGS}}\rangle$.
Their projector product corresponds to the projector $|S_{\text{SGS}}\rangle\langle S_{\text{SGS}}|$.

To analyze the effect of an additional centralizer constraint, let $P\in {\Gamma}(H)$ be a Hermitian Pauli operator and suppose
\begin{equation}
    P|E_0\rangle=p|E_0\rangle, \ p\in\{\pm1\}.
\end{equation}
This condition holds automatically for a nondegenerate ground state. 
Then the corresponding projector is
\begin{equation}
    \Pi_s=\frac{I+sP}{2}.
\end{equation}
By applying the projector on the target state as 
\begin{align}
    \Pi_s|E_0\rangle=\left\{
   \begin{array}{ll}
     |E_0\rangle,\  s=p,  \\[4pt]
     0,\ s=-p. 
\end{array}
\right.\label{proj}
\end{align}
For a normalized input state $|\psi\rangle$, let
\begin{equation}
    |\psi_s\rangle=\frac{\Pi_s|\psi\rangle}{\sqrt{q_s}}.
\end{equation}
For a matching sign $s=p$, its fidelity satisfies
\begin{align}
    {\cal F}(|E_0\rangle,|\psi_p\rangle)&=\frac{|\langle E_0|\Pi_p|\psi\rangle|^2}{\langle \psi|\Pi_p|\psi\rangle} \nonumber\\
    &=\frac{{\cal F}(|E_0\rangle,|\psi\rangle)}{q_p} \nonumber\\
    &\geq {\cal F}(|E_0\rangle,|\psi\rangle).
\end{align}
For $0<p_q\leq 1$. For the opposite sign, the projected state has zero overlap with $|E_0\rangle$ which will indicate that the normalized state has $0$ fidelity.
We note that for the degenerate Hamiltonian case, $P$ with appropriate signs will project the states within the ground state eigenspace which will not effect the ground state fidelity.

\section{\MakeUppercase{Complexity analysis of the algorithm}}\label{appenc}

To isolate the effect of initialization, we analyze a sequence of consecutive $M_0$ outcomes without interventing unitary corrections.
We assume a nondegenerate ground state and a normalized initial state $|\psi_{\text{OSGS}}\rangle$ with nonzero ground state overlap. 
We define 
\begin{equation}
    f_n= \mathcal{F}(|E_n\rangle,|\psi_{\text{OSGS}}\rangle),\ a_n=\cos^2(\epsilon E_n+\frac{\pi}{4}).
\end{equation}
The probability of obtaining $k$ consecutive $M_0$ outcomes is 
\begin{equation}
    S_k=\sum_n f_n (a_n)^k,\ S_0=1.
\end{equation}
Conditioned on this sequence, the normalized state is
\begin{equation}
    |\psi_k\rangle=\frac{M_0^k|\psi_{OSGS}\rangle}{\sqrt{S_k}},
\end{equation}
For this sequence, its ground state fidelity then is 
\begin{align}
    \mathcal{F}(|E_0\rangle,|\psi_{k}\rangle)&=\frac{f_0(a_0)^k}{S_k} \nonumber \\
    &=\big[1+\sum_{n>0}\frac{f_n}{f_0}(\frac{a_n}{a_0})^k\big]^{-1}.\label{fm0kps}
\end{align}
Since $\epsilon>0$ and $\epsilon E_n\in [-\pi/4,\pi/4]$, $a_n$  strictly decreases with energy. 
Thus, $a_n<a_0$ for every occupied excited level with $E_n>E_0$, and $\mathcal{F}(|E_0\rangle,|\psi_{k}\rangle)$ increases toward unity provided that $f_0>0$.
Let $E_1>E_0$ be the lowest excited state energy and define $a_1=\cos^2(\epsilon E_1+\pi/4)$.
Then $a_n\leq a_1$ for all excited states, whether or not the first excited level is occupied initially.
In the case of $0<f_0<1$ and $0<a_1<a_0$, for a given algorithm accuracy $\delta$, a sufficient measurement count to achieve $1-\mathcal{F}(|E_0\rangle,|\psi_{k}\rangle)\leq\delta$ is 
\begin{equation}
    k_\delta=\max\left\{0,\ \left\lceil
\frac{\ln\!\left[\dfrac{(1-f_0)(1-\delta)}{f_0\delta}\right]}
{\ln(a_0/a_1)}
\right\rceil\right\}.\label{meatimesande0}
\end{equation}
If $f_0=1$, no filtering is required. If all occupied excited state value satisfy $a_n=0$, one $M_0$ measurement outcome removes these components.

This expression characterizes the conditional filtering sequence, not the total cost of obtaining it.
To illustrate the distinction, consider an idealized restart protocol that terminates an attempt at the first $M_1$ outcome, reprepares the same initial state, and declares success after $k$ consecutive $M_0$ outcomes.

The probability of failure at step $t$ is $S_{t-1}-S_t$, and the expected number of measurements in one attempt is 
\begin{equation}
    L_k=\sum_{t=1}^k t(S_{t-1}-S_t)+kS_k =\sum_{j=0}^{k-1}S_j.
\end{equation}
Since each attempt succeeds with probability $S_k$, the expectated total measurement count is 
\begin{equation}
    \mathbb{E}[T]=\frac{\sum_{j=0}^{k-1}S_j}{S_k}.
\end{equation}
Since $S_k\geq f_0 a_0^k$ and $\sum_{j=0}^{k-1}S_j\leq k$, choosing $k=k_{\delta}$ yields
\begin{equation}
    \mathbb{E}[T] \leq\frac{k_{\delta}}{f_0 a_0^{k_{\delta}}}, \label{ttot}
\end{equation}
where $k_\delta$ is sufficient to achieve a conditional ground state fidelity of at least $1-\delta$.
This bound accounts for both the successful measurement sequence and the overhead from failed attempts.

Equation ~(\ref{ttot}) bounds the expected number of weak measurements required by the idealized restart protocol to achieve the target conditional fidelity.
The bound depends on the initial ground state weight, the measurement factor $a_0$ and $a_1$ and the target infidelity $\delta$.
A polynomial scaling claim requires additional assumptions on how these quantities vary with system size.

\section{\MakeUppercase{Algorithm error estimation}}\label{append}

To access the conditional filtering error, we consider a measurement sequence whose amplitude peak lies at the acceptance boundary
\begin{equation}
 x_{k_0 k_1}^{\max}=\epsilon E_{\text{th}},   
\end{equation} 
where the proposed threshold choice is $E_{\text{th}}=E_{\text{SGS}}$.

Let the initial state be $|\psi_{\text{OSGS}}\rangle$.
After $K=k_0+k_1$  weak measurements without unitary corrections, the normalized resulting state is given by Eq.~(\ref{psik}).

With the assumption of sufficiently weak measurements ($\epsilon\ll1$) and a sufficiently large total number of measurements $K=k_0+k_1$, the amplitude function can be approximated by a Gaussian distribution. Specifically, $f_K(x)$ can be expressed as
\begin{equation}
    f_K(x)\approx f_K(x_{k_0 k_1}^{\max})e^{-K(x-x_{k_0k_1}^{\max})^2},
\end{equation}
This approximation is valid at the energy levels relevant to the following error estimate.

To quantify the performance of weak-measurement process, we evaluate the normalized fidelity between the normalized post-measurement state and true ground state after $K=k_0+k_1$ weak measurements 
\begin{align}
    \mathcal{F}(|E_0\rangle,|\psi_K\rangle)&=\frac{|\widetilde{A}_0^{(K)}|^2}{\sum_n|\widetilde{A}_n^{(K)}|^2}\nonumber\\ 
    &=\big[1+\sum_{n>0}\big|\frac{A_nf_K(\epsilon E_n)}{A_0f_K(\epsilon E_0)}\big|^2\big]^{-1}.
\end{align}
The initial state satisfies $A_0\neq0$.

Since the amplitude function $A(\epsilon E_n)$ decays rapidly for $E_n>E_0$ when the measurement time increases, the contribution from higher-energy states becomes negligible, such that $f_K(\epsilon E_1)>f_K(\epsilon E_n)$ for $ n\geq2$.
The fidelity can be expressed as
\begin{align}
    \mathcal{F}(|E_0\rangle,|\psi_K\rangle) &\leq \big[1+\big|\frac{A_1 f_K(\epsilon E_1)}{A_0f_K(\epsilon E_0)}\big|^2\big]^{-1} \nonumber\\ 
    &\leq 1-|\frac{A_1f_K(\epsilon E_1)}{A_0f_K(\epsilon E_0)}|^2. \label{fapprox}
\end{align}
For the  degenerate energy case, the total initial population should replace the single population in these expressions.

This approximation becomes increasingly accurate for large $K$, where the weak-measurement process strongly suppresses the excited-state individual, yielding a fidelity that approaches unity.

The conditional filtering error is defined as
\begin{equation}
    \mathcal{E}_K=1-\mathcal{F}(|E_0\rangle,|\psi_K\rangle).
\end{equation}
Substituting Eq.~(\ref{fapprox}) into the above definition yields 
\begin{equation}
    \mathcal{E}_K\geq \frac{\mathcal{F}(|\psi_{\text{OSGS}}\rangle,|E_1\rangle)}{\mathcal{F}(|\psi_{\text{OSGS}}\rangle,|E_0\rangle)}e^{2K\epsilon^2(E_0-E_1)(E_0+E_1-2E_{\text{th}})}.
\end{equation}
This expression shows that the conditional error is governed by three factors: the initial excited-to-ground fidelity ratio, the number of measurements $K$, and the position of the energy threshold respect to the first two energy levels. 
With this formula, the exponential factor decreases with $K$ when
\begin{equation}
    E_{\text{th}}<\frac{E_0+E_1}{2}.
\end{equation}
Under these conditions, an OSGS initialization that reduces the initial excited-to-ground fidelity ratio reduces the leading error prefactor and can reduce the number of measurements required to reach a prescribed fidelity. 

\end{appendices}



\bibliography{reference}

@article{PhysRevLett.118.090501,
  title = {Application of a Resource Theory for Magic States to Fault-Tolerant Quantum Computing},
  author = {Howard, Mark and Campbell, Earl},
  journal = {Phys. Rev. Lett.},
  volume = {118},
  issue = {9},
  pages = {090501},
  numpages = {6},
  year = {2017},
  month = {Mar},
  publisher = {American Physical Society}
}

@article{crawford2021efficient,
  title={Efficient quantum measurement of Pauli operators in the presence of finite sampling error},
  author={Crawford, Ophelia and Van Straaten, Barnaby and Wang, Daochen and Parks, Thomas and Campbell, Earl and Brierley, Stephen},
  journal={Quantum},
  volume={5},
  pages={385},
  year={2021},
  publisher={Verein zur F{\"o}rderung des Open Access Publizierens in den Quantenwissenschaften}
}

@article{sun2025stabilizer,
  title={Stabilizer ground states for simulating quantum many-body physics: theory, algorithms, and applications},
  author={Sun, Jiace and Cheng, Lixue and Zhang, Shi-Xin},
  journal={Quantum},
  volume={9},
  pages={1782},
  year={2025},
  publisher={Verein zur F{\"o}rderung des Open Access Publizierens in den Quantenwissenschaften}
}

@article{sarkar2021sets,
  title={On sets of maximally commuting and anticommuting Pauli operators},
  author={Sarkar, Rahul and van den Berg, Ewout},
  journal={Research in the Mathematical Sciences},
  volume={8},
  number={1},
  pages={14},
  year={2021},
  publisher={Springer}
}

@article{mao2023measurement,
  title={Measurement-based deterministic imaginary time evolution},
  author={Mao, Yuping and Chaudhary, Manish and Kondappan, Manikandan and Shi, Junheng and Ilo-Okeke, Ebubechukwu O and Ivannikov, Valentin and Byrnes, Tim},
  journal={Physical Review Letters},
  volume={131},
  number={11},
  pages={110602},
  year={2023},
  publisher={APS}
}

@article{steane1998quantum,
  title={Quantum computing},
  author={Steane, Andrew},
  journal={Reports on Progress in Physics},
  volume={61},
  number={2},
  pages={117},
  year={1998},
  publisher={IOP Publishing}
}

@article{georgescu2014quantum,
  title={Quantum simulation},
  author={Georgescu, Iulia M and Ashhab, Sahel and Nori, Franco},
  journal={Reviews of Modern Physics},
  volume={86},
  number={1},
  pages={153--185},
  year={2014},
  publisher={APS}
}

@article{bloch2008many,
  title={Many-body physics with ultracold gases},
  author={Bloch, Immanuel and Dalibard, Jean and Zwerger, Wilhelm},
  journal={Reviews of modern physics},
  volume={80},
  number={3},
  pages={885--964},
  year={2008},
  publisher={APS}
}

@book{levine2009quantum,
  title={Quantum chemistry},
  author={Levine, Ira N and Busch, Daryle H and Shull, Harrison},
  volume={6},
  year={2009},
  publisher={Pearson Prentice Hall Upper Saddle River, NJ}
}

@article{byrnes2007quantum,
  title={Quantum Simulator for the Hubbard Model with Long-Range Coulomb Interactions Using Surface Acoustic Waves},
  author={Byrnes, Tim and Recher, Patrik and Kim, Na Young and Utsunomiya, Shoko and Yamamoto, Yoshihisa},
  journal={Physical review letters},
  volume={99},
  number={1},
  pages={016405},
  year={2007},
  publisher={APS}
}

@article{lloyd1996universal,
  title={Universal quantum simulators},
  author={Lloyd, Seth},
  journal={Science},
  volume={273},
  number={5278},
  pages={1073--1078},
  year={1996},
  publisher={American Association for the Advancement of Science}
}

@article{bluvstein2024logical,
  title={Logical quantum processor based on reconfigurable atom arrays},
  author={Bluvstein, Dolev and Evered, Simon J and Geim, Alexandra A and Li, Sophie H and Zhou, Hengyun and Manovitz, Tom and Ebadi, Sepehr and Cain, Madelyn and Kalinowski, Marcin and Hangleiter, Dominik and others},
  journal={Nature},
  volume={626},
  number={7997},
  pages={58--65},
  year={2024},
  publisher={Nature Publishing Group UK London}
}

@article{arute2019quantum,
  title={Quantum supremacy using a programmable superconducting processor},
  author={Arute, Frank and Arya, Kunal and Babbush, Ryan and Bacon, Dave and Bardin, Joseph C and Barends, Rami and Biswas, Rupak and Boixo, Sergio and Brandao, Fernando GSL and Buell, David A and others},
  journal={Nature},
  volume={574},
  number={7779},
  pages={505--510},
  year={2019},
  publisher={Nature Publishing Group UK London}
}

@book{williams2010explorations,
  title={Explorations in quantum computing},
  author={Williams, Colin P},
  year={2010},
  publisher={Springer Science \& Business Media}
}

@book{tasaki2020physics,
  title={Physics and mathematics of quantum many-body systems},
  author={Tasaki, Hal},
  volume={66},
  year={2020},
  publisher={Springer}
}

@article{vidal2004efficient,
  title={Efficient simulation of one-dimensional quantum many-body systems},
  author={Vidal, Guifr{\'e}},
  journal={Physical review letters},
  volume={93},
  number={4},
  pages={040502},
  year={2004},
  publisher={APS}
}

@article{eisert2015quantum,
  title={Quantum many-body systems out of equilibrium},
  author={Eisert, Jens and Friesdorf, Mathis and Gogolin, Christian},
  journal={Nature Physics},
  volume={11},
  number={2},
  pages={124--130},
  year={2015},
  publisher={Nature Publishing Group UK London}
}

@article{daley2022practical,
  title={Practical quantum advantage in quantum simulation},
  author={Daley, Andrew J and Bloch, Immanuel and Kokail, Christian and Flannigan, Stuart and Pearson, Natalie and Troyer, Matthias and Zoller, Peter},
  journal={Nature},
  volume={607},
  number={7920},
  pages={667--676},
  year={2022},
  publisher={Nature Publishing Group UK London}
}

@article{bauer2023quantum,
  title={Quantum simulation for high-energy physics},
  author={Bauer, Christian W and Davoudi, Zohreh and Balantekin, A Baha and Bhattacharya, Tanmoy and Carena, Marcela and De Jong, Wibe A and Draper, Patrick and El-Khadra, Aida and Gemelke, Nate and Hanada, Masanori and others},
  journal={PRX quantum},
  volume={4},
  number={2},
  pages={027001},
  year={2023},
  publisher={APS}
}

@article{nachman2021quantum,
  title={Quantum algorithm for high energy physics simulations},
  author={Nachman, Benjamin and Provasoli, Davide and De Jong, Wibe A and Bauer, Christian W},
  journal={Physical review letters},
  volume={126},
  number={6},
  pages={062001},
  year={2021},
  publisher={APS}
}

@article{hofstetter2018quantum,
  title={Quantum simulation of strongly correlated condensed matter systems},
  author={Hofstetter, Walter and Qin, Tao},
  journal={Journal of Physics B: Atomic, Molecular and Optical Physics},
  volume={51},
  number={8},
  pages={082001},
  year={2018},
  publisher={IOP Publishing}
}

@article{dovesi2018quantum,
  title={Quantum-mechanical condensed matter simulations with CRYSTAL},
  author={Dovesi, Roberto and Erba, Alessandro and Orlando, Roberto and Zicovich-Wilson, Claudio M and Civalleri, Bartolomeo and Maschio, Lorenzo and R{\'e}rat, Michel and Casassa, Silvia and Baima, Jacopo and Salustro, Simone and others},
  journal={Wiley Interdisciplinary Reviews: Computational Molecular Science},
  volume={8},
  number={4},
  pages={e1360},
  year={2018},
  publisher={Wiley Online Library}
}

@article{ge2019faster,
  title={Faster ground state preparation and high-precision ground energy estimation with fewer qubits},
  author={Ge, Yimin and Tura, Jordi and Cirac, J Ignacio},
  journal={Journal of Mathematical Physics},
  volume={60},
  pages = {022202},
  year={2019},
  publisher={AIP Publishing}
}

@inproceedings{gleinig2021efficient,
  title={An efficient algorithm for sparse quantum state preparation},
  author={Gleinig, Niels and Hoefler, Torsten},
  booktitle={2021 58th ACM/IEEE Design Automation Conference (DAC)},
  pages={433--438},
  year={2021},
  organization={IEEE}
}

@article{berry2025rapid,
  title={Rapid Initial-State Preparation for the Quantum Simulation of Strongly Correlated Molecules},
  author={Berry, Dominic W and Tong, Yu and Khattar, Tanuj and White, Alec and Kim, Tae In and Low, Guang Hao and Boixo, Sergio and Ding, Zhiyan and Lin, Lin and Lee, Seunghoon and others},
  journal={PRX Quantum},
  volume={6},
  number={2},
  pages={020327},
  year={2025},
  publisher={APS}
}

@inproceedings{aharonov2003adiabatic,
  title={Adiabatic quantum state generation and statistical zero knowledge},
  author={Aharonov, Dorit and Ta-Shma, Amnon},
  booktitle={Proceedings of the thirty-fifth annual ACM symposium on Theory of computing},
  pages={20--29},
  year={2003}
}

@article{abrams1997simulation,
  title={Simulation of many-body Fermi systems on a universal quantum computer},
  author={Abrams, Daniel S and Lloyd, Seth},
  journal={Physical Review Letters},
  volume={79},
  number={13},
  pages={2586},
  year={1997},
  publisher={APS}
}

@article{wang2009efficient,
  title={Efficient quantum algorithm for preparing molecular-system-like states on a quantum computer},
  author={Wang, Hefeng and Ashhab, Sahel and Nori, Franco},
  journal={Physical Review A—Atomic, Molecular, and Optical Physics},
  volume={79},
  number={4},
  pages={042335},
  year={2009},
  publisher={APS}
}

@article{chiu2018quantum,
  title={Quantum state engineering of a Hubbard system with ultracold fermions},
  author={Chiu, Christie S and Ji, Geoffrey and Mazurenko, Anton and Greif, Daniel and Greiner, Markus},
  journal={Physical review letters},
  volume={120},
  number={24},
  pages={243201},
  year={2018},
  publisher={APS}
}

@article{muller2007state,
  title={State preparation and dynamics of ultracold atoms in higher lattice orbitals},
  author={M{\"u}ller, Torben and F{\"o}lling, Simon and Widera, Artur and Bloch, Immanuel},
  journal={Physical Review Letters},
  volume={99},
  number={20},
  pages={200405},
  year={2007},
  publisher={APS}
}

@article{tarruell2018quantum,
  title={Quantum simulation of the Hubbard model with ultracold fermions in optical lattices},
  author={Tarruell, Leticia and Sanchez-Palencia, Laurent},
  journal={Comptes Rendus Physique},
  volume={19},
  number={6},
  pages={365--393},
  year={2018},
  publisher={Elsevier}
}

@article{hensgens2017quantum,
  title={Quantum simulation of a Fermi--Hubbard model using a semiconductor quantum dot array},
  author={Hensgens, Toivo and Fujita, Takafumi and Janssen, Laurens and Li, Xiao and Van Diepen, CJ and Reichl, Christian and Wegscheider, Werner and Das Sarma, Sankar and Vandersypen, Lieven MK},
  journal={Nature},
  volume={548},
  number={7665},
  pages={70--73},
  year={2017},
  publisher={Nature Publishing Group UK London}
}

@article{huggins2025efficient,
  title={Efficient state preparation for the quantum simulation of molecules in first quantization},
  author={Huggins, William J and Leimkuhler, Oskar and Stetina, Torin F and Whaley, K Birgitta},
  journal={PRX Quantum},
  volume={6},
  number={2},
  pages={020319},
  year={2025},
  publisher={APS}
}

@article{long2001efficient,
  title={Efficient scheme for initializing a quantum register with an arbitrary superposed state},
  author={Long, Gui-Lu and Sun, Yang},
  journal={Physical Review A},
  volume={64},
  number={1},
  pages={014303},
  year={2001},
  publisher={APS}
}

@article{aspuru2005simulated,
  title={Simulated quantum computation of molecular energies},
  author={Aspuru-Guzik, Al{\'a}n and Dutoi, Anthony D and Love, Peter J and Head-Gordon, Martin},
  journal={Science},
  volume={309},
  number={5741},
  pages={1704--1707},
  year={2005},
  publisher={American Association for the Advancement of Science}
}

@article{suzuki1993improved,
  title={Improved Trotter-like formula},
  author={Suzuki, Masuo},
  journal={Physics Letters A},
  volume={180},
  number={3},
  pages={232--234},
  year={1993},
  publisher={Elsevier}
}

@book{gottesman1997stabilizer,
  title={Stabilizer codes and quantum error correction},
  author={Gottesman, Daniel},
  year={1997},
  publisher={California Institute of Technology}
}

@article{gottesman2009introduction,
  title={An introduction to quantum error correction and fault-tolerant quantum computation},
  author={Gottesman, Daniel},
  journal={arXiv preprint arXiv:0904.2557},
  year={2009}
}

@article{kitaev1997quantum,
  title={Quantum computations: algorithms and error correction},
  author={Kitaev, A Yu},
  journal={Russian Mathematical Surveys},
  volume={52},
  number={6},
  pages={1191},
  year={1997},
  publisher={IOP Publishing}
}

@article{laflamme1996perfect,
  title={Perfect quantum error correcting code},
  author={Laflamme, Raymond and Miquel, Cesar and Paz, Juan Pablo and Zurek, Wojciech Hubert},
  journal={Physical Review Letters},
  volume={77},
  number={1},
  pages={198},
  year={1996},
  publisher={APS}
}

@article{gottesman1996class,
  title={Class of quantum error-correcting codes saturating the quantum Hamming bound},
  author={Gottesman, Daniel},
  journal={Physical Review A},
  volume={54},
  number={3},
  pages={1862},
  year={1996},
  publisher={APS}
}

@article{trotter1959product,
  title={On the product of semi-groups of operators},
  author={Trotter, Hale F},
  journal={Proceedings of the American Mathematical Society},
  volume={10},
  number={4},
  pages={545--551},
  year={1959},
  publisher={JSTOR}
}

@article{poulin2009preparing,
  title={Preparing ground states of quantum many-body systems on a quantum computer},
  author={Poulin, David and Wocjan, Pawel},
  journal={Physical review letters},
  volume={102},
  number={13},
  pages={130503},
  year={2009},
  publisher={APS}
}

@article{fauseweh2024quantum,
  title={Quantum many-body simulations on digital quantum computers: State-of-the-art and future challenges},
  author={Fauseweh, Benedikt},
  journal={Nature Communications},
  volume={15},
  number={1},
  pages={2123},
  year={2024},
  publisher={Nature Publishing Group UK London}
}

@article{kapit2012non,
  title={Non-Abelian braiding of lattice bosons},
  author={Kapit, Eliot and Ginsparg, Paul and Mueller, Erich},
  journal={Physical Review Letters},
  volume={108},
  number={6},
  pages={066802},
  year={2012},
  publisher={APS}
}

@article{kandala2017hardware,
  title={Hardware-efficient variational quantum eigensolver for small molecules and quantum magnets},
  author={Kandala, Abhinav and Mezzacapo, Antonio and Temme, Kristan and Takita, Maika and Brink, Markus and Chow, Jerry M and Gambetta, Jay M},
  journal={nature},
  volume={549},
  number={7671},
  pages={242--246},
  year={2017},
  publisher={Nature Publishing Group}
}

@article{stannigel2014constrained,
  title={Constrained dynamics via the Zeno effect in quantum simulation: Implementing non-Abelian lattice gauge theories with cold atoms},
  author={Stannigel, K and Hauke, Philipp and Marcos, David and Hafezi, Mohammad and Diehl, S and Dalmonte, M and Zoller, P},
  journal={Physical review letters},
  volume={112},
  number={12},
  pages={120406},
  year={2014},
  publisher={APS}
}

@article{gottesman1998heisenberg,
  title={The Heisenberg representation of quantum computers},
  author={Gottesman, Daniel},
  journal={arXiv preprint quant-ph/9807006},
  year={1998}
}

@article{kueng2017low,
  title={Low rank matrix recovery from rank one measurements},
  author={Kueng, Richard and Rauhut, Holger and Terstiege, Ulrich},
  journal={Applied and Computational Harmonic Analysis},
  volume={42},
  number={1},
  pages={88--116},
  year={2017},
  publisher={Elsevier}
}

@article{flammia2012quantum,
  title={Quantum tomography via compressed sensing: error bounds, sample complexity and efficient estimators},
  author={Flammia, Steven T and Gross, David and Liu, Yi-Kai and Eisert, Jens},
  journal={New Journal of Physics},
  volume={14},
  number={9},
  pages={095022},
  year={2012},
  publisher={IOP Publishing}
}

@article{gross2010quantum,
  title={Quantum state tomography via compressed sensing},
  author={Gross, David and Liu, Yi-Kai and Flammia, Steven T and Becker, Stephen and Eisert, Jens},
  journal={Physical review letters},
  volume={105},
  number={15},
  pages={150401},
  year={2010},
  publisher={APS}
}

@article{cerezo2022variational,
  title={Variational quantum state eigensolver},
  author={Cerezo, Marco and Sharma, Kunal and Arrasmith, Andrew and Coles, Patrick J},
  journal={npj Quantum Information},
  volume={8},
  number={1},
  pages={113},
  year={2022},
  publisher={Nature Publishing Group UK London}
}

@article{cleve1998quantum,
  title={Quantum algorithms revisited},
  author={Cleve, Richard and Ekert, Artur and Macchiavello, Chiara and Mosca, Michele},
  journal={Proceedings of the Royal Society of London. Series A: Mathematical, Physical and Engineering Sciences},
  volume={454},
  number={1969},
  pages={339--354},
  year={1998},
  publisher={The Royal Society}
}

@book{nielsen2010quantum,
  title={Quantum computation and quantum information},
  author={Nielsen, Michael A and Chuang, Isaac L},
  year={2010},
  publisher={Cambridge university press}
}

@article{aaronson2004improved,
  title={Improved simulation of stabilizer circuits},
  author={Aaronson, Scott and Gottesman, Daniel},
  journal={Physical Review A—Atomic, Molecular, and Optical Physics},
  volume={70},
  number={5},
  pages={052328},
  year={2004},
  publisher={APS}
}

@article{wu2015review,
  title={A review on algorithms for maximum clique problems},
  author={Wu, Qinghua and Hao, Jin-Kao},
  journal={European Journal of Operational Research},
  volume={242},
  number={3},
  pages={693--709},
  year={2015},
  publisher={Elsevier}
}

@article{feige2004approximating,
  title={Approximating maximum clique by removing subgraphs},
  author={Feige, Uriel},
  journal={SIAM Journal on Discrete Mathematics},
  volume={18},
  number={2},
  pages={219--225},
  year={2004},
  publisher={SIAM}
}

@book{butenko2003maximum,
  title={Maximum independent set and related problems, with applications},
  author={Butenko, Sergiy},
  year={2003},
  publisher={University of Florida}
}

@article{wu2013adaptive,
  title={An adaptive multistart tabu search approach to solve the maximum clique problem},
  author={Wu, Qinghua and Hao, Jin-Kao},
  journal={Journal of Combinatorial Optimization},
  volume={26},
  number={1},
  pages={86--108},
  year={2013},
  publisher={Springer}
}

@article{guturu2008impatient,
  title={An impatient evolutionary algorithm with probabilistic tabu search for unified solution of some NP-hard problems in graph and set theory via clique finding},
  author={Guturu, Parthasarathy and Dantu, Ram},
  journal={IEEE Transactions on Systems, Man, and Cybernetics, Part B (Cybernetics)},
  volume={38},
  number={3},
  pages={645--666},
  year={2008},
  publisher={IEEE}
}

@inproceedings{chapuis2017finding,
  title={Finding maximum cliques on a quantum annealer},
  author={Chapuis, Guillaume and Djidjev, Hristo and Hahn, Georg and Rizk, Guillaume},
  booktitle={Proceedings of the Computing Frontiers Conference},
  pages={63--70},
  year={2017}
}

@article{cheng2011finding,
  title={Finding maximal cliques in massive networks},
  author={Cheng, James and Ke, Yiping and Fu, Ada Wai-Chee and Yu, Jeffrey Xu and Zhu, Linhong},
  journal={ACM Transactions on Database Systems (TODS)},
  volume={36},
  number={4},
  pages={21},
  year={2011},
  publisher={ACM New York, NY, USA}
}

@article{herrman2021impact,
  title={Impact of graph structures for QAOA on MaxCut},
  author={Herrman, Rebekah and Treffert, Lorna and Ostrowski, James and Lotshaw, Phillip C and Humble, Travis S and Siopsis, George},
  journal={Quantum Information Processing},
  volume={20},
  number={9},
  pages={289},
  year={2021},
  publisher={Springer}
}

@article{gidney2021stim,
  title={Stim: a fast stabilizer circuit simulator},
  author={Gidney, Craig},
  journal={Quantum},
  volume={5},
  pages={497},
  year={2021},
  publisher={Verein zur F{\"o}rderung des Open Access Publizierens in den Quantenwissenschaften}
}

@article{masot2024stabilizer,
  title={Stabilizer tensor networks: Universal quantum simulator on a basis of stabilizer states},
  author={Masot-Llima, Sergi and Garcia-Saez, Artur},
  journal={Physical Review Letters},
  volume={133},
  number={23},
  pages={230601},
  year={2024},
  publisher={APS}
}

@article{de2011linearized,
  title={A linearized stabilizer formalism for systems of finite dimension},
  author={De Beaudrap, Niel},
  journal={Quantum Information and Computation},
  volume={13},
  pages={73-115},
  year={2013}
}

@article{byrnes2006simulating,
  title={Simulating lattice gauge theories on a quantum computer},
  author={Byrnes, Tim and Yamamoto, Yoshihisa},
  journal={Physical Review A—Atomic, Molecular, and Optical Physics},
  volume={73},
  number={2},
  pages={022328},
  year={2006},
  publisher={APS}
}

@article{heyl2013dynamical,
  title={Dynamical quantum phase transitions in the transverse-field Ising model},
  author={Heyl, Markus and Polkovnikov, Anatoli and Kehrein, Stefan},
  journal={Physical review letters},
  volume={110},
  number={13},
  pages={135704},
  year={2013},
  publisher={APS}
}

@article{mohseni2022ising,
  title={Ising machines as hardware solvers of combinatorial optimization problems},
  author={Mohseni, Naeimeh and McMahon, Peter L and Byrnes, Tim},
  journal={Nature Reviews Physics},
  volume={4},
  number={6},
  pages={363--379},
  year={2022},
  publisher={Nature Publishing Group UK London}
}

@article{bennink2017unbiased,
  title={Unbiased simulation of near-Clifford quantum circuits},
  author={Bennink, Ryan S and Ferragut, Erik M and Humble, Travis S and Laska, Jason A and Nutaro, James J and Pleszkoch, Mark G and Pooser, Raphael C},
  journal={Physical Review A},
  volume={95},
  number={6},
  pages={062337},
  year={2017},
  publisher={APS}
}

@phdthesis{leonetti2025deterministic,
  title={A Deterministic Decoration Scheme for Measurement-Based Variational Quantum Eigensolvers},
  author={Leonetti, Martina},
  year={2025},
  school={Politecnico di Torino}
}

@book{byrnes2021quantum,
  title={Quantum atom optics: Theory and applications to quantum technology},
  author={Byrnes, Tim and Ilo-Okeke, Ebubechukwu O},
  year={2021},
  publisher={Cambridge university press}
}

@article{ladd2010quantum,
  title={Quantum computers},
  author={Ladd, Thaddeus D and Jelezko, Fedor and Laflamme, Raymond and Nakamura, Yasunobu and Monroe, Christopher and O’Brien, Jeremy Lloyd},
  journal={nature},
  volume={464},
  number={7285},
  pages={45--53},
  year={2010},
  publisher={Nature Publishing Group UK London}
}

@article{hamaguchi2024handbook,
  title={Handbook for quantifying robustness of magic},
  author={Hamaguchi, Hiroki and Hamada, Kou and Yoshioka, Nobuyuki},
  journal={Quantum},
  volume={8},
  pages={1461},
  year={2024},
  publisher={Verein zur F{\"o}rderung des Open Access Publizierens in den Quantenwissenschaften}
}

@article{gyongyosi2019survey,
  title={A survey on quantum computing technology},
  author={Gyongyosi, Laszlo and Imre, Sandor},
  journal={Computer Science Review},
  volume={31},
  pages={51--71},
  year={2019},
  publisher={Elsevier}
}

@article{robin2025stabilizer,
  title={Stabilizer-accelerated quantum many-body ground-state estimation},
  author={Robin, Caroline EP},
  journal={Physical Review A},
  volume={112},
  number={5},
  pages={052408},
  year={2025},
  publisher={APS}
}

@article{dehaene2003clifford,
  title={Clifford group, stabilizer states, and linear and quadratic operations over GF (2)},
  author={Dehaene, Jeroen and De Moor, Bart},
  journal={Physical Review A},
  volume={68},
  number={4},
  pages={042318},
  year={2003},
  publisher={APS}
}

@article{calderbank1998quantum,
  title={Quantum error correction via codes over GF (4)},
  author={Calderbank, A Robert and Rains, Eric M and Shor, Peter M and Sloane, Neil JA},
  journal={IEEE Transactions on Information Theory},
  volume={44},
  number={4},
  pages={1369--1387},
  year={1998},
  publisher={IEEE}
}

@article{wilde2009logical,
  title={Logical operators of quantum codes},
  author={Wilde, Mark M},
  journal={Physical Review A—Atomic, Molecular, and Optical Physics},
  volume={79},
  number={6},
  pages={062322},
  year={2009},
  publisher={APS}
}

@article{kirby2020classical,
  title={Classical simulation of noncontextual Pauli Hamiltonians},
  author={Kirby, William M and Love, Peter J},
  journal = {Phys. Rev. A},
  volume = {102},
  number = {13},
  pages = {032418},
  year={2020},
  publisher={AIP Publishing}
}

@article{johan2001some,
  title={Some optimal inapproximability results},
  author={H{\aa}stad, Johan},
  journal={Journal of the ACM},
  volume={48},
  number={4},
  pages={498--859},
  year={2001}
}

@article{berlekamp1978inherent,
  title={On the inherent intractability of certain coding problems (corresp.)},
  author={Berlekamp, Elwyn and McEliece, Robert and Van Tilborg, Henk},
  journal={IEEE Transactions on Information theory},
  volume={24},
  number={3},
  pages={384--386},
  year={1978},
  publisher={IEEE}
}

@article{ermakov2026unified,
  title={Unified framework for efficiently computable quantum circuits},
  author={Ermakov, Igor and Lychkovskiy, Oleg and Byrnes, Tim},
  journal={EPJ Quantum Technology},
  year={2026},
  publisher={Springer}
}
\end{document}